\documentclass[reprint,twocolumn,superscriptaddress,showpacs]{revtex4}
\usepackage{graphicx}
\usepackage[varg]{txfonts}
\usepackage{dcolumn}
\usepackage{bm}
\usepackage{color}
\usepackage{latexsym}
\usepackage{supertabular}
\usepackage{hyperref}
\usepackage{longtable}

\begin{document}

\hyphenpenalty=5000
\doublehyphendemerits=5000
\finalhyphendemerits=5000

\title{Benchmark calculations for electron-impact excitation of Mg$^{4+}$}

\author{K.~Wang}
\affiliation{College of Physics and Materials Science, Henan Normal University, Xinxiang, 453007, P.R.~China}
\affiliation{Department of Physics and Astronomy, Drake University, Des Moines, Iowa, 50311, USA}
\author{L.~Fern\'{a}ndez-Menchero}
\author{O.~Zatsarinny}
\author{K.~Bartschat}
\affiliation{Department of Physics and Astronomy, Drake University, Des Moines, Iowa, 50311, USA}

\date{\today}
\begin{abstract}
There are major discrepancies between recent \hbox{B-spline} \hbox{R-matrix} (BSR) and
Dirac Atomic \hbox{R-matrix} Code (DARC) calculations regarding electron-impact excitation rates
for transitions in $\mathrm{Mg}^{4+}$, with claims that the DARC calculations are much more accurate.
To identify possible reasons for these discrepancies and to estimate the accuracy
of the various results, we carried out independent BSR calculations
with the same 86 target states as in the previous calculations,
but with a different and more accurate representation of the target structure.
We find close agreement with the previous BSR results for the majority of transitions,
thereby confirming their accuracy. At the same time the differences with the DARC
results are much more pronounced.
The discrepancies in the final results for the collision strengths are mainly due to differences
in the structure description, specifically the inclusion of correlation effects, and due to the likely 
occurrence of pseudo\-resonances.  
To further check the convergence of the predicted collision rates, we carried out even more
extensive calculations involving 316 states of $\mathrm{Mg}^{4+}$.
Extending the close-coupling expansion results in major corrections for transitions involving
the higher-lying states
and allows us to assess the likely uncertainties in the existing datasets.
\end{abstract}


\maketitle

\section{\label{sec:Introduction} Introduction}

Accurate and reliable electron-impact excitation rates and transition probabilities
are required for the modeling and spectroscopic diagnostics of various non\-equilibrium
astrophysical and laboratory plasmas. As a common situation for most atomic ions, there is a limited
number of measurements available (if there are any at all) for the transition probabilities and
especially for the electron-impact excitation rates. Consequently, theoretical predictions
are used in most applications.
Despite the enormous progress made during the past three decades both in the theory and
the computational methods of treating electron collisions with atoms, ions, and molecules,
the calculation of collision rates remains a serious challenge
for complex many-electron targets. The accuracy of the existing datasets is not well established,
and often there are situations when subsequent, more extended calculations suggest considerable
corrections to previous results. Even when employing similar scattering models,
calculations with different methods or codes may also lead to large discrepancies.
A recent example of such a situation is electron-impact excitation of $\mathrm{Mg}^{4+}$,
which is the subject of the present paper.

Spectral lines of $\mathrm{Mg}^{4+}$ have been observed in several astro\-physical plasmas, including the
Sun. Due to their sensitivity to the density and temperature, these lines are used
in the diagnostics of such plasmas.
Many references and examples are given in the recent publications
by Hudson and co-workers~\cite{hudson2009b}, Tayal and Sossah~\cite{tayal2015}, 
and Aggarwal and Keenan~\cite{aggarwal2016b}.
All these works employed the advanced \hbox{R-matrix} (close-coupling) method for the collision process,
but different computer codes and different representations of the target structure.
Hudson {\it et al.}~\cite{hudson2009b} reported effective collision strengths for
transitions between 37 fine-structure levels of the $\mathrm{2s^2\,2p^4}$, $\mathrm{2s\,2p^5}$,
$\mathrm{2p^6}$, $\mathrm{2s^2\,2p^3\,3s}$, and $\mathrm{2s^2\,2p^3\,3p}$ configurations.
[We omit listing the closed $\rm (1s^2)$ sub\-shell here and below.]
The target wave\-functions were first obtained in the non\-relativistic approximation
for 19 $LS$-coupled target states using the CIV3 code~\cite{hibbert1975}.
The configuration interaction (CI) expansion included up to 1350 configurations and $4l$ correlated orbitals.
The \hbox{RMATRX-1} codes \cite{berrington1995} were then employed for the $LS$ scattering calculations,
and the Intermediate Coupling Frame Transformation (ICFT) method~\cite{griffin1998}
was applied to generate the intermediate-coupling results.
These calculations will be denoted as \hbox{RM-37} below.
Hudson~{\it et al.} concluded that whilst the overall accuracy is difficult to assess,
they expected their results to have an accuracy of $\pm 10\%$. As will be seen below by comparison
with results from other calculations, these
uncertainty estimates were most likely far too optimistic.

To begin with, the work of Hudson and co-workers~\cite{hudson2009b} may contain some uncertainties
due to the omission of the levels of the $\mathrm{2s^2\,2p^3\,3d}$ configuration
in the $n=3$ complex. Tayal and Sossah~\cite{tayal2015}, therefore, performed a more extensive calculation for
\hbox{$\mathrm{e}-\mathrm{Mg}^{4+}$} collisions.
They additionally took into consideration the levels of the $\mathrm{2s^2\,2p^3\,3d}$ configuration, 
resulting overall in 86 fine-structure levels. 
They also improved the target structure description further by employing
the extensive multi\-configuration Hartree-Fock (MCHF) method \cite{froese-fischer2007} in combination with
term-dependent non\-orthogonal orbitals, which were individually optimized for the various $LS$ terms.
The scattering calculations for the collision rates were performed with the \hbox{B-spline} \hbox{R-matrix} (BSR)
method (see~\cite{zatsarinny2013} for an overview) and a parallelized version of the associated computer 
code~\cite{zatsarinny2006} in the semi-relativistic Breit-Pauli approximation.
One advantage of this approach is the avoidance of pseudo\-resonances, which were possibly
faced by Hudson and Bell~\cite{hudson2006}
in their standard \hbox{R-matrix} calculations with orthogonal orbitals.
The calculation of Tayal and Sossah~\cite{tayal2015} will be denoted as  \hbox{BSR-86 (TS)} below.

The effective collision strengths were presented over a wide range of electron temperatures,
suitable for modeling the emissions from various types of astrophysical plasmas.
Tayal and Sossah~\cite{tayal2015} estimated the accuracy of their collision rates as $\sim \!10\%$ for
transitions from the $\mathrm{2s^2\,2p^4}$ levels of the ground-state configuration and somewhat
less accurate ($\sim \! 20\%$) for transitions between excited levels.
For higher levels of the $\mathrm{2s^2\,2p^3\,3d}$ configuration in particular, there
may be significant
coupling effects from higher states, and thus their results for transitions involving
these levels are likely less accurate ($\sim\!25\%$).

Recently, Aggarwal and Keenan~\cite{aggarwal2016b} noted major differences between the \hbox{RM-37}
and  \hbox{BSR-86 (TS)}  results.
They estimated that the  \hbox{BSR-86 (TS)}  and \hbox{RM-37}  collision strengths differ for about
$80\%$ of the transitions by more than $20\%$,
and in most cases the  \hbox{BSR-86 (TS)}  results are larger.
Aggarwal and Keenan~\cite{aggarwal2016b} suspected (with reasons discussed below in
Sec.~\ref{sec:results}) major inaccuracies of the BSR collision strengths,
and hence they performed another calculation for \hbox{$\mathrm{e}-\mathrm{Mg}^{4+}$} collisions.
They included the same 86 levels as in the  \hbox{BSR-86 (TS)}  model,
but they employed a completely different, fully relativistic approach.
The target wave\-functions used by Aggarwal and Keenan~\cite{aggarwal2016b} were generated with the
General-purpose Relativistic Atomic Structure Package (GRASP) code based on the $jj$-coupling scheme.
The target configuration expansions included 12 configurations,
namely $\mathrm{2s^2\,2p^4}$, $\mathrm{2s\,2p^5}$, $\mathrm{2p^6}$,
$\mathrm{2s^2\,2p^3\,3}l$, $\mathrm{2s\,2p^4\,3}l$, and $\mathrm{2p^5\,3}l$, respectively.
This is less than in the CI expansions of the target states in the \hbox{RM-37}
calculations and certainly in the \hbox{BSR-86 (TS)} model.
Due to the limited CI included in the model,
the target description is less accurate than in the  \hbox{BSR-86 (TS)}  and \hbox{RM-37}  models.
The subsequent scattering calculations were performed with the Dirac atomic \hbox{R-matrix} code (DARC),
which also includes relativistic effects in a systematic way based on
the $jj$-coupling scheme.
Both the atomic structure (GRASP) and scattering (DARC) codes are available at the website
\url{http://www.apap-network.org/codes.html}.
Although they adopted
fully relativistic codes for their calculations of both the ionic structure and
the collision parameters, Aggarwal and Keenan~\cite{aggarwal2016b} stressed the well-known fact that relativistic effects
are not expected to be very important in the predicted collisions rates for such a comparatively light
ion as $\mathrm{Mg}^{4+}$. We will denote their calculation by \hbox{DARC-86}.

The comparison between the final collision strengths show significant discrepancies
(up to two orders of magnitude) between the \hbox{DARC-86}  and  \hbox{BSR-86 (TS)}  predictions
for over $60\%$ of the transitions at all temperatures.
Aggarwal and Keenan~\cite{aggarwal2016b} stated that the scale of the discrepancies cannot be explained
by differences in the atomic structure alone,
and therefore the  \hbox{BSR-86 (TS)}  results should be classified as inaccurate.
They also noticed that the BSR collision strengths seem to overestimate the results at higher energies
and exhibit incorrect trends regarding the temperature dependence of the predicted collision rates.
As a possible reason, they suggested the appearance of pseudo\-resonances.
Finally, they concluded that their \hbox{DARC-86}  collision rates are probably the best available
to date and should be adopted for the modeling and diagnostics of plasmas.

The principal motivation for the present work, therefore, was to shed more light
on the ongoing discussion by performing an independent calculation for
electron collisions with $\mathrm{Mg}^{4+}$ and thereby to respond to recent demands
for uncertainty estimates of theoretical predictions~\cite{PhysRevA.83.040001,chung2016}.
We also use the BSR approach, but with an entirely different target description,
again based on a \hbox{B-spline} expansion.
By comparing with other available, and presumably highly accurate, structure-only calculations, we
are confident that the present
target description is significantly more accurate than any of those used before in scattering calculations.
For a proper comparison of the results, we select the same set of 86 target states in
the close-coupling (CC) expansion as in previous  \hbox{BSR-86 (TS)}  and \hbox{DARC-86} calculations.
Our model will be labeled~\hbox{BSR-86}.
Choosing the same target states allows us to directly draw conclusions regarding the sensitivity of
the predictions to the target structure description in the cases of interest.

As a second goal, we wanted to check the convergence of the CC expansion, especially for
transitions to higher-lying states.  To achieve this, we performed much more extensive calculations,
also including
the levels of the $\mathrm{2s^2\,2p^4\,4}l$ and $\mathrm{2s\,2p^5\,3}l$ configurations for a total of 316 coupled states.
This model will be referred to as \hbox{BSR-316}.
Note that this extension allows us to consider the important $2\mathrm{s}-3l$ and $3l-4l$ transitions, which
may modify the close-coupling effects and change the background collision strengths.
Even more important, the inclusion of additional resonance series converging to higher excited
levels can change the effective collision strengths significantly for
the $\mathrm{Mg}^{4+}$ ion around the temperature of its formation in solar plasmas.
Note that the maximum $\mathrm{Mg}^{4+}$ abundance in ionization equilibrium of
solar plasmas occurs around
$\mathrm{T_e} = 2.8 \times 10^5\,\mathrm{K}$~\cite{mazzotta1998}.

This paper is organized as follows.  We begin in Sec.~\ref{sec:theory} by summarizing
the most important features of the present BSR models for the $\mathrm{e}-\mathrm{Mg}^{4+}$ scattering process.
This is followed in Sec.~\ref{sec:results} with a presentation and discussion of our present results,
in comparison with those from previous calculations.
Then we present our most extensive model including 316 target states.
We finish with a brief summary and conclusions in Sec.~\ref{sec:summary}.
Unless specified otherwise, atomic units are used throughout.

\section{\label{sec:theory} Computational details} 

\subsection{\label{subsec:struct} Structure calculations} 

The target states of $\mathrm{Mg}^{4+}$ in the present calculations were generated by combining
the MCHF and the \hbox{B-spline} box-based multi\-channel methods \cite{zatsarinny2009}.
Specifically, the structure of the target expansion was chosen as
\begin{eqnarray}\label{eq:expn}
\Phi(2s^22p^3 nl, LSJ)
   =  \sum_{nl,L'S'} a_{nl}^{L'S'LSJ} \left\{\phi(2s^22p^3,L'S')P_{nl}\right\}^{LSJ} \nonumber\\
   +  \sum_{nl,L'S'} b_{nl}^{L'S'LSJ} \left\{\phi(2s2p^4,L'S')P_{nl}\right\}^{LSJ} \nonumber\\
  +  c^{\,LSJ}  \varphi(2s^22p^4)^{LSJ} + d^{\,LSJ}  \varphi(2s2p^5)^{LSJ} + e^{\,LSJ}  \varphi(2p^6)^{LSJ}. 
  \nonumber\\ \hspace{-10.0truemm}
\end{eqnarray}
Here $P_{nl}$ denotes the orbital of the outer valence electron,
while the $\phi$ and $\varphi$ functions represent the CI expansions
of the corresponding ionic or specific atomic states, respectively. These expansions
were generated in separate MCHF calculations for each state.

The expansion (\ref{eq:expn}) can be considered as a model for the entire $\mathrm{2s^2\,2p^3}\,nl$
and $\mathrm{2s\,2p^4}\,nl$ Rydberg series of bound and auto\-ionizing states in O-like Mg$^{4+}$.
The two sums in this expansion can also provide a good approximation for states with equivalent
electrons, namely for all terms of the ground-state configuration $\mathrm{2s^2\,2p^4}$ as well as
for the core-excited states $\mathrm{2s\,2p^5}$.
We found, however, that it is more appropriate to employ separate CI expansions for these states
by directly including relaxation and term-dependence effects generated with state-specific one-electron orbitals.

The inner-core (short-range) correlation is accounted for through the CI expansion of the
$\mathrm{2s^2\,2p^3}$ and $\mathrm{2s\,2p^4}$ ionic states.
These expansions include all single and double excitations
from the $\mathrm{2s}$ and $\mathrm{2p}$ orbitals to the $3l$ and $4l$ \hbox{($l = 0-3$)}
correlated orbitals.
The resulting ionization potentials for all ionic states agreed with
the NIST tables~\cite{nist2016} within $0.01\,\mathrm{eV}$.
To keep the final expansions for the target states to a manageable size,
all CI expansions were restricted by dropping contributions with coefficients
whose magnitude was less than the cut-off parameter of $0.025$.
Even though this is a significant reduction of the configuration expansions,
we had to compromise in order to make the subsequent scattering calculations
feasible on the available computers.

The unknown functions $P_{nl}$ for the outer valence electron were expanded in a
B-spline basis, and the corresponding equations were solved subject to the
condition that the orbitals vanish at the boundary.
The \hbox{R-matrix} radius was set to $12\,a_0$,
where $a_0 = 0.529 \times 10^{-10}\,\mathrm{m}$ is the Bohr radius.
We employed 82 \hbox{B-splines} of order~8 to span this radial range using a semi-exponential knot grid.
The \hbox{B-spline} coefficients for the valence electron orbitals $P_{nl}$, along with the CI coefficients
$a_{nl}^{L'S'LSJ}$, $b_{nl}^{L'S'LSJ}$, $c^{\,LSJ}$, $d^{\,LSJ}$, and $e^{\,LSJ}$ in Eq.~(\ref{eq:expn}),
were obtained by a variational method through diagonalizing the target Hamiltonian
in the Breit-Pauli approximation.
Relativistic effects  were incorporated
through the Darwin, mass velocity, and one-electron spin-orbit operators.
This is sufficient to account for the main relativistic corrections for light ions such as Mg$^{4+}$.
Since the \hbox{B-spline} bound-state close-coupling calculations generate different
non\-orthogonal sets of orbitals for each ionic state, their subsequent use is somewhat
complicated. Our configuration expansions for the target states
contained from 20 to at most 50 configurations for each state.
Such expansions can be used in collision calculations
with modern computational resources.

\begin{longtable*}{rllrr@{\ }rr@{\ }rr@{\ }rr@{\ }r}
\hline\hline
 $  i$ &              Configuration           &                Level &       NIST &       BSR  & $  \% $ &      MCHF  & $  \% $ & BSR-86 (TS)& $  \% $ &    DARC-86 & $  \% $ \\
\hline
\endfirsthead

\hline\hline
 $  i$ &              Configuration           &                Level &       NIST &       BSR  & $  \% $ &      MCHF  & $  \% $ & BSR-86 (TS)& $  \% $ &   DARC-86 & $  \% $ \\
\hline
\endhead

\hline
\hline
\endfoot

\hline
\hline
\caption{Excitation energies of the 86 target levels included in the present BSR scattering calculations \\
Key: $i$: level index;
NIST: recommended value from NIST data base~\cite{nist2016};
BSR: present calculation, 
MCHF: calculation~\cite{froese-fischer2004};
\hbox{BSR-86 (TS)}: calculation~\cite{tayal2015};
DARC-86: calculation~\cite{aggarwal2016b};
$\%$: deviation respect the recommended values of NIST, in percentage.
All energies in $\mathrm{cm}^{-1}$.} \label{tab:energies} 
\endlastfoot

 $  1$ & $\mathrm{2s^2\,2p^4}               $ & $\mathrm{^3P_2    }$ & $     0.0$ & $     0.0$ & $   - $ & $     0.0$ & $   - $ & $     0.0$ & $   - $ & $     0.0$ & $   - $ \\
 $  2$ & $\mathrm{2s^2\,2p^4}               $ & $\mathrm{^3P_1    }$ & $  1782.1$ & $  1610.6$ & $ -9.6$ & $  1778.2$ & $ -0.2$ & $  1729.5$ & $ -3.0$ & $  1733.8$ & $ -2.7$ \\
 $  3$ & $\mathrm{2s^2\,2p^4}               $ & $\mathrm{^3P_0    }$ & $  2520.7$ & $  2558.2$ & $  1.5$ & $  2506.8$ & $ -0.6$ & $  2507.5$ & $ -0.5$ & $  2454.8$ & $ -2.6$ \\
 $  4$ & $\mathrm{2s^2\,2p^4}               $ & $\mathrm{^1D_2    }$ & $ 35924.7$ & $ 35519.1$ & $ -1.1$ & $ 36207.2$ & $  0.8$ & $ 35968.6$ & $  0.1$ & $ 39198.2$ & $  9.1$ \\
 $  5$ & $\mathrm{2s^2\,2p^4}               $ & $\mathrm{^1S_0    }$ & $ 77286.9$ & $ 77497.5$ & $  0.3$ & $ 77291.2$ & $  0.0$ & $ 77192.5$ & $ -0.1$ & $ 77508.6$ & $  0.3$ \\
 $  6$ & $\mathrm{2s  \,2p^5}               $ & $\mathrm{^3P_2^{o}}$ & $283211.2$ & $283191.8$ & $ -0.0$ & $283885.0$ & $  0.2$ & $284222.9$ & $  0.4$ & $293811.8$ & $  3.7$ \\
 $  7$ & $\mathrm{2s  \,2p^5}               $ & $\mathrm{^3P_1^{o}}$ & $284830.9$ & $284846.2$ & $  0.0$ & $285496.0$ & $  0.2$ & $285820.7$ & $  0.3$ & $295374.4$ & $  3.7$ \\
 $  8$ & $\mathrm{2s  \,2p^5}               $ & $\mathrm{^3P_0^{o}}$ & $285714.3$ & $285691.8$ & $ -0.0$ & $286374.9$ & $  0.2$ & $286636.1$ & $  0.3$ & $296222.7$ & $  3.7$ \\
 $  9$ & $\mathrm{2s  \,2p^5}               $ & $\mathrm{^1P_1^{o}}$ & $397485.0$ & $396900.1$ & $ -0.1$ & $399069.2$ & $  0.4$ & $400600.5$ & $  0.8$ & $420037.1$ & $  5.7$ \\
 $ 10$ & $\mathrm{2p^6}                     $ & $\mathrm{^1S_0    }$ & $662938.5$ & $662251.9$ & $ -0.1$ & $665826.2$ & $  0.4$ & $665680.8$ & $  0.4$ & $703031.0$ & $  6.0$ \\
 $ 11$ & $\mathrm{2s^2\,2p^3\,(^4S_3)\,3s  }$ & $\mathrm{^5S_2^{o}}$ & $668933.4$ & $672099.1$ & $  0.5$ & $668750.8$ & $ -0.0$ & $669132.1$ & $  0.0$ & $654687.3$ & $ -2.1$ \\
 $ 12$ & $\mathrm{2s^2\,2p^3\,(^4S_3)\,3s  }$ & $\mathrm{^3S_1^{o}}$ & $684540.3$ & $687940.3$ & $  0.5$ & $684358.8$ & $ -0.0$ & $684092.5$ & $ -0.1$ & $671168.8$ & $ -2.0$ \\
 $ 13$ & $\mathrm{2s^2\,2p^3\,(^2D_3)\,3s  }$ & $\mathrm{^3D_1^{o}}$ & $727781.2$ & $730753.6$ & $  0.4$ & $727999.5$ & $  0.0$ & $727455.2$ & $ -0.0$ & $717109.2$ & $ -1.5$ \\
 $ 14$ & $\mathrm{2s^2\,2p^3\,(^2D_3)\,3s  }$ & $\mathrm{^3D_2^{o}}$ & $727762.5$ & $730790.2$ & $  0.4$ & $727977.5$ & $  0.0$ & $727424.5$ & $ -0.0$ & $717086.2$ & $ -1.5$ \\
 $ 15$ & $\mathrm{2s^2\,2p^3\,(^2D_3)\,3s  }$ & $\mathrm{^3D_3^{o}}$ & $727741.7$ & $730879.0$ & $  0.4$ & $727961.2$ & $  0.0$ & $727400.4$ & $ -0.0$ & $717081.8$ & $ -1.5$ \\
 $ 16$ & $\mathrm{2s^2\,2p^3\,(^4S_3)\,3p  }$ & $\mathrm{^5P_1    }$ & $      - $ & $735026.0$ & $   - $ & $731732.9$ & $   - $ & $731793.2$ & $   - $ & $717840.1$ & $   - $ \\
 $ 17$ & $\mathrm{2s^2\,2p^3\,(^4S_3)\,3p  }$ & $\mathrm{^5P_2    }$ & $      - $ & $735147.3$ & $   - $ & $731854.4$ & $   - $ & $731919.4$ & $   - $ & $717960.8$ & $   - $ \\
 $ 18$ & $\mathrm{2s^2\,2p^3\,(^4S_3)\,3p  }$ & $\mathrm{^5P_3    }$ & $      - $ & $735335.1$ & $   - $ & $732056.8$ & $   - $ & $732118.0$ & $   - $ & $718163.8$ & $   - $ \\
 $ 19$ & $\mathrm{2s^2\,2p^3\,(^2D_3)\,3s  }$ & $\mathrm{^1D_2^{o}}$ & $735545.1$ & $738799.9$ & $  0.4$ & $735852.3$ & $  0.0$ & $735155.5$ & $ -0.1$ & $725280.3$ & $ -1.4$ \\
 $ 20$ & $\mathrm{2s^2\,2p^3\,(^4S_3)\,3p  }$ & $\mathrm{^3P_1    }$ & $      - $ & $750115.3$ & $   - $ & $745983.5$ & $   - $ & $745769.3$ & $   - $ & $734490.5$ & $   - $ \\
 $ 21$ & $\mathrm{2s^2\,2p^3\,(^4S_3)\,3p  }$ & $\mathrm{^3P_0    }$ & $      - $ & $750144.4$ & $   - $ & $746042.8$ & $   - $ & $745807.7$ & $   - $ & $734544.3$ & $   - $ \\
 $ 22$ & $\mathrm{2s^2\,2p^3\,(^4S_3)\,3p  }$ & $\mathrm{^3P_2    }$ & $      - $ & $750172.7$ & $   - $ & $746017.7$ & $   - $ & $745803.3$ & $   - $ & $734554.2$ & $   - $ \\
 $ 23$ & $\mathrm{2s^2\,2p^3\,(^2P_1)\,3s  }$ & $\mathrm{^3P_0^{o}}$ & $756544.4$ & $759814.0$ & $  0.4$ & $756663.3$ & $  0.0$ & $757755.9$ & $  0.2$ & $742034.9$ & $ -1.9$ \\
 $ 24$ & $\mathrm{2s^2\,2p^3\,(^2P_1)\,3s  }$ & $\mathrm{^3P_1^{o}}$ & $756565.3$ & $759856.6$ & $  0.4$ & $756691.0$ & $  0.0$ & $757791.0$ & $  0.2$ & $742071.2$ & $ -1.9$ \\
 $ 25$ & $\mathrm{2s^2\,2p^3\,(^2P_1)\,3s  }$ & $\mathrm{^3P_2^{o}}$ & $756641.0$ & $759955.8$ & $  0.4$ & $756777.1$ & $  0.0$ & $757872.2$ & $  0.2$ & $742180.9$ & $ -1.9$ \\
 $ 26$ & $\mathrm{2s^2\,2p^3\,(^2P_1)\,3s  }$ & $\mathrm{^1P_1^{o}}$ & $764627.7$ & $768204.8$ & $  0.5$ & $764879.0$ & $  0.0$ & $764823.0$ & $  0.0$ & $750664.7$ & $ -1.8$ \\
 $ 27$ & $\mathrm{2s^2\,2p^3\,(^2D_3)\,3p  }$ & $\mathrm{^1P_1    }$ & $      - $ & $787177.7$ & $   - $ & $784332.3$ & $   - $ & $784291.5$ & $   - $ & $773595.4$ & $   - $ \\
 $ 28$ & $\mathrm{2s^2\,2p^3\,(^2D_3)\,3p  }$ & $\mathrm{^3D_1    }$ & $      - $ & $789183.0$ & $   - $ & $786679.6$ & $   - $ & $786654.1$ & $   - $ & $775674.9$ & $   - $ \\
 $ 29$ & $\mathrm{2s^2\,2p^3\,(^2D_3)\,3p  }$ & $\mathrm{^3D_2    }$ & $      - $ & $789185.5$ & $   - $ & $786675.2$ & $   - $ & $786647.6$ & $   - $ & $775633.2$ & $   - $ \\
 $ 30$ & $\mathrm{2s^2\,2p^3\,(^2D_3)\,3p  }$ & $\mathrm{^3D_3    }$ & $      - $ & $789574.6$ & $   - $ & $786958.4$ & $   - $ & $786915.3$ & $   - $ & $775948.2$ & $   - $ \\
 $ 31$ & $\mathrm{2s^2\,2p^3\,(^2D_3)\,3p  }$ & $\mathrm{^3F_2    }$ & $      - $ & $793827.8$ & $   - $ & $791237.2$ & $   - $ & $791392.6$ & $   - $ & $780107.2$ & $   - $ \\
 $ 32$ & $\mathrm{2s^2\,2p^3\,(^2D_3)\,3p  }$ & $\mathrm{^3F_3    }$ & $      - $ & $794002.8$ & $   - $ & $791353.4$ & $   - $ & $791499.0$ & $   - $ & $780227.9$ & $   - $ \\
 $ 33$ & $\mathrm{2s^2\,2p^3\,(^2D_3)\,3p  }$ & $\mathrm{^3F_4    }$ & $      - $ & $794219.1$ & $   - $ & $791500.2$ & $   - $ & $791642.8$ & $   - $ & $780377.2$ & $   - $ \\
 $ 34$ & $\mathrm{2s^2\,2p^3\,(^2D_3)\,3p  }$ & $\mathrm{^1F_3    }$ & $      - $ & $796987.4$ & $   - $ & $793929.3$ & $   - $ & $794104.2$ & $   - $ & $782923.1$ & $   - $ \\
 $ 35$ & $\mathrm{2s^2\,2p^3\,(^2D_3)\,3p  }$ & $\mathrm{^3P_1    }$ & $      - $ & $810674.4$ & $   - $ & $807494.9$ & $   - $ & $808699.3$ & $   - $ & $799153.2$ & $   - $ \\
 $ 36$ & $\mathrm{2s^2\,2p^3\,(^2D_3)\,3p  }$ & $\mathrm{^3P_0    }$ & $      - $ & $810760.9$ & $   - $ & $807610.0$ & $   - $ & $808816.7$ & $   - $ & $800241.8$ & $   - $ \\
 $ 37$ & $\mathrm{2s^2\,2p^3\,(^2D_3)\,3p  }$ & $\mathrm{^3P_2    }$ & $      - $ & $810779.1$ & $   - $ & $807482.2$ & $   - $ & $808827.7$ & $   - $ & $800898.0$ & $   - $ \\
 $ 38$ & $\mathrm{2s^2\,2p^3\,(^4S_3)\,3d  }$ & $\mathrm{^5D_0^{o}}$ & $      - $ & $814649.7$ & $   - $ & $811397.1$ & $   - $ & $811460.3$ & $   - $ & $797638.8$ & $   - $ \\
 $ 39$ & $\mathrm{2s^2\,2p^3\,(^4S_3)\,3d  }$ & $\mathrm{^5D_1^{o}}$ & $      - $ & $814656.8$ & $   - $ & $811393.1$ & $   - $ & $811460.3$ & $   - $ & $797633.4$ & $   - $ \\
 $ 40$ & $\mathrm{2s^2\,2p^3\,(^4S_3)\,3d  }$ & $\mathrm{^5D_2^{o}}$ & $      - $ & $814671.1$ & $   - $ & $811386.0$ & $   - $ & $811458.1$ & $   - $ & $797624.6$ & $   - $ \\
 $ 41$ & $\mathrm{2s^2\,2p^3\,(^4S_3)\,3d  }$ & $\mathrm{^5D_3^{o}}$ & $      - $ & $814692.3$ & $   - $ & $811377.8$ & $   - $ & $811455.9$ & $   - $ & $797613.6$ & $   - $ \\
 $ 42$ & $\mathrm{2s^2\,2p^3\,(^4S_3)\,3d  }$ & $\mathrm{^5D_4^{o}}$ & $      - $ & $814721.0$ & $   - $ & $811372.6$ & $   - $ & $811450.4$ & $   - $ & $797607.0$ & $   - $ \\
 $ 43$ & $\mathrm{2s^2\,2p^3\,(^2P_1)\,3p  }$ & $\mathrm{^3S_1    }$ & $      - $ & $817811.7$ & $   - $ & $814970.6$ & $   - $ & $814948.8$ & $   - $ & $801496.1$ & $   - $ \\
 $ 44$ & $\mathrm{2s^2\,2p^3\,(^2D_3)\,3p  }$ & $\mathrm{^1D_2    }$ & $      - $ & $821273.8$ & $   - $ & $817471.1$ & $   - $ & $817522.1$ & $   - $ & $804903.5$ & $   - $ \\
 $ 45$ & $\mathrm{2s^2\,2p^3\,(^2P_1)\,3p  }$ & $\mathrm{^3D_3    }$ & $      - $ & $822737.4$ & $   - $ & $819668.0$ & $   - $ & $820259.0$ & $   - $ & $805064.8$ & $   - $ \\
 $ 46$ & $\mathrm{2s^2\,2p^3\,(^2P_1)\,3p  }$ & $\mathrm{^3D_1    }$ & $      - $ & $822738.8$ & $   - $ & $819630.2$ & $   - $ & $820204.1$ & $   - $ & $805061.5$ & $   - $ \\
 $ 47$ & $\mathrm{2s^2\,2p^3\,(^2P_1)\,3p  }$ & $\mathrm{^3D_2    }$ & $      - $ & $823003.7$ & $   - $ & $819841.6$ & $   - $ & $820381.9$ & $   - $ & $807325.4$ & $   - $ \\
 $ 48$ & $\mathrm{2s^2\,2p^3\,(^4S_3)\,3d  }$ & $\mathrm{^3D_1^{o}}$ & $821973.1$ & $825791.2$ & $  0.5$ & $821989.5$ & $  0.0$ & $821609.9$ & $ -0.0$ & $810405.7$ & $ -1.4$ \\
 $ 49$ & $\mathrm{2s^2\,2p^3\,(^4S_3)\,3d  }$ & $\mathrm{^3D_2^{o}}$ & $821988.5$ & $825818.4$ & $  0.5$ & $822006.4$ & $  0.0$ & $821632.9$ & $ -0.0$ & $810428.7$ & $ -1.4$ \\
 $ 50$ & $\mathrm{2s^2\,2p^3\,(^4S_3)\,3d  }$ & $\mathrm{^3D_3^{o}}$ & $822065.3$ & $825904.6$ & $  0.5$ & $822090.0$ & $  0.0$ & $821710.8$ & $ -0.0$ & $810527.5$ & $ -1.4$ \\
 $ 51$ & $\mathrm{2s^2\,2p^3\,(^2P_1)\,3p  }$ & $\mathrm{^1P_1    }$ & $      - $ & $826740.9$ & $   - $ & $823564.3$ & $   - $ & $824355.5$ & $   - $ & $809697.9$ & $   - $ \\
 $ 52$ & $\mathrm{2s^2\,2p^3\,(^2P_1)\,3p  }$ & $\mathrm{^3P_2    }$ & $      - $ & $831883.1$ & $   - $ & $828259.1$ & $   - $ & $829754.6$ & $   - $ & $815304.4$ & $   - $ \\
 $ 53$ & $\mathrm{2s^2\,2p^3\,(^2P_1)\,3p  }$ & $\mathrm{^3P_1    }$ & $      - $ & $831928.9$ & $   - $ & $828312.4$ & $   - $ & $829937.8$ & $   - $ & $815939.7$ & $   - $ \\
 $ 54$ & $\mathrm{2s^2\,2p^3\,(^2P_1)\,3p  }$ & $\mathrm{^3P_0    }$ & $      - $ & $831960.0$ & $   - $ & $828329.8$ & $   - $ & $830031.1$ & $   - $ & $816244.8$ & $   - $ \\
 $ 55$ & $\mathrm{2s^2\,2p^3\,(^2P_1)\,3p  }$ & $\mathrm{^1D_2    }$ & $      - $ & $840273.8$ & $   - $ & $836464.9$ & $   - $ & $836690.0$ & $   - $ & $831570.7$ & $   - $ \\
 $ 56$ & $\mathrm{2s^2\,2p^3\,(^2P_1)\,3p  }$ & $\mathrm{^1S_0    }$ & $      - $ & $864130.1$ & $   - $ & $859521.5$ & $   - $ & $860403.1$ & $   - $ & $857826.5$ & $   - $ \\
 $ 57$ & $\mathrm{2s^2\,2p^3\,(^2D_3)\,3d  }$ & $\mathrm{^3F_2^{o}}$ & $      - $ & $868272.9$ & $   - $ & $865979.7$ & $   - $ & $866135.8$ & $   - $ & $854888.8$ & $   - $ \\
 $ 58$ & $\mathrm{2s^2\,2p^3\,(^2D_3)\,3d  }$ & $\mathrm{^3F_3^{o}}$ & $      - $ & $868583.7$ & $   - $ & $866201.0$ & $   - $ & $866350.9$ & $   - $ & $855147.8$ & $   - $ \\
 $ 59$ & $\mathrm{2s^2\,2p^3\,(^2D_3)\,3d  }$ & $\mathrm{^3F_4^{o}}$ & $      - $ & $868986.1$ & $   - $ & $866485.0$ & $   - $ & $866617.5$ & $   - $ & $855473.7$ & $   - $ \\
 $ 60$ & $\mathrm{2s^2\,2p^3\,(^2D_3)\,3d  }$ & $\mathrm{^1S_0^{o}}$ & $      - $ & $869176.8$ & $   - $ & $866323.2$ & $   - $ & $866623.0$ & $   - $ & $855387.0$ & $   - $ \\
 $ 61$ & $\mathrm{2s^2\,2p^3\,(^2D_3)\,3d  }$ & $\mathrm{^3G_3^{o}}$ & $      - $ & $871964.8$ & $   - $ & $869482.5$ & $   - $ & $870045.7$ & $   - $ & $858809.7$ & $   - $ \\
 $ 62$ & $\mathrm{2s^2\,2p^3\,(^2D_3)\,3d  }$ & $\mathrm{^3G_4^{o}}$ & $      - $ & $872044.1$ & $   - $ & $869466.7$ & $   - $ & $870024.9$ & $   - $ & $858794.4$ & $   - $ \\
 $ 63$ & $\mathrm{2s^2\,2p^3\,(^2D_3)\,3d  }$ & $\mathrm{^3G_5^{o}}$ & $      - $ & $872124.9$ & $   - $ & $869433.3$ & $   - $ & $869989.8$ & $   - $ & $858762.5$ & $   - $ \\
 $ 64$ & $\mathrm{2s^2\,2p^3\,(^2D_3)\,3d  }$ & $\mathrm{^1G_4^{o}}$ & $      - $ & $873523.7$ & $   - $ & $870354.2$ & $   - $ & $870823.8$ & $   - $ & $860180.3$ & $   - $ \\
 $ 65$ & $\mathrm{2s^2\,2p^3\,(^2D_3)\,3d  }$ & $\mathrm{^3D_3^{o}}$ & $871215.5$ & $874276.4$ & $  0.4$ & $871754.2$ & $  0.1$ & $872364.5$ & $  0.1$ & $861522.4$ & $ -1.1$ \\
 $ 66$ & $\mathrm{2s^2\,2p^3\,(^2D_3)\,3d  }$ & $\mathrm{^3D_2^{o}}$ & $871357.1$ & $874370.0$ & $  0.3$ & $871888.8$ & $  0.1$ & $872467.6$ & $  0.1$ & $861597.0$ & $ -1.1$ \\
 $ 67$ & $\mathrm{2s^2\,2p^3\,(^2D_3)\,3d  }$ & $\mathrm{^3D_1^{o}}$ & $871390.0$ & $874383.4$ & $  0.3$ & $871931.8$ & $  0.1$ & $872481.9$ & $  0.1$ & $861586.1$ & $ -1.1$ \\
 $ 68$ & $\mathrm{2s^2\,2p^3\,(^2D_3)\,3d  }$ & $\mathrm{^1P_1^{o}}$ & $873455.3$ & $876697.1$ & $  0.4$ & $874012.1$ & $  0.1$ & $873427.8$ & $ -0.0$ & $863719.4$ & $ -1.1$ \\
 $ 69$ & $\mathrm{2s^2\,2p^3\,(^2D_3)\,3d  }$ & $\mathrm{^3P_2^{o}}$ & $876794.6$ & $879669.2$ & $  0.3$ & $877334.9$ & $  0.1$ & $877491.4$ & $  0.1$ & $867044.4$ & $ -1.1$ \\
 $ 70$ & $\mathrm{2s^2\,2p^3\,(^2D_3)\,3d  }$ & $\mathrm{^3P_1^{o}}$ & $877282.9$ & $880189.1$ & $  0.3$ & $877817.2$ & $  0.1$ & $877935.8$ & $  0.1$ & $867692.9$ & $ -1.1$ \\
 $ 71$ & $\mathrm{2s^2\,2p^3\,(^2D_3)\,3d  }$ & $\mathrm{^3P_0^{o}}$ & $877462.9$ & $880370.0$ & $  0.3$ & $877996.2$ & $  0.1$ & $878118.0$ & $  0.1$ & $867925.6$ & $ -1.1$ \\
 $ 72$ & $\mathrm{2s^2\,2p^3\,(^2D_3)\,3d  }$ & $\mathrm{^1D_2^{o}}$ & $877611.0$ & $881148.5$ & $  0.4$ & $878202.1$ & $  0.1$ & $878639.2$ & $  0.1$ & $867169.5$ & $ -1.2$ \\
 $ 73$ & $\mathrm{2s^2\,2p^3\,(^2D_3)\,3d  }$ & $\mathrm{^3S_1^{o}}$ & $879515.0$ & $882786.3$ & $  0.4$ & $880101.8$ & $  0.1$ & $879580.8$ & $  0.0$ & $868757.4$ & $ -1.2$ \\
 $ 74$ & $\mathrm{2s^2\,2p^3\,(^2D_3)\,3d  }$ & $\mathrm{^1F_3^{o}}$ & $882790.6$ & $886303.1$ & $  0.4$ & $883467.4$ & $  0.1$ & $883872.6$ & $  0.1$ & $875261.5$ & $ -0.9$ \\
 $ 75$ & $\mathrm{2s^2\,2p^3\,(^2P_1)\,3d  }$ & $\mathrm{^3F_4^{o}}$ & $      - $ & $901002.3$ & $   - $ & $897995.0$ & $   - $ & $899251.2$ & $   - $ & $883480.9$ & $   - $ \\
 $ 76$ & $\mathrm{2s^2\,2p^3\,(^2P_1)\,3d  }$ & $\mathrm{^3F_3^{o}}$ & $      - $ & $901241.0$ & $   - $ & $898266.7$ & $   - $ & $899511.3$ & $   - $ & $883803.5$ & $   - $ \\
 $ 77$ & $\mathrm{2s^2\,2p^3\,(^2P_1)\,3d  }$ & $\mathrm{^3F_2^{o}}$ & $      - $ & $901388.5$ & $   - $ & $898441.8$ & $   - $ & $899673.7$ & $   - $ & $883990.0$ & $   - $ \\
 $ 78$ & $\mathrm{2s^2\,2p^3\,(^2P_1)\,3d  }$ & $\mathrm{^3P_0^{o}}$ & $898756.3$ & $902183.6$ & $  0.4$ & $899165.8$ & $  0.0$ & $900107.2$ & $  0.2$ & $884033.9$ & $ -1.6$ \\
 $ 79$ & $\mathrm{2s^2\,2p^3\,(^2P_1)\,3d  }$ & $\mathrm{^3P_1^{o}}$ & $898961.5$ & $902449.9$ & $  0.4$ & $899374.5$ & $  0.0$ & $900308.0$ & $  0.1$ & $884292.9$ & $ -1.6$ \\
 $ 80$ & $\mathrm{2s^2\,2p^3\,(^2P_1)\,3d  }$ & $\mathrm{^3P_2^{o}}$ & $899368.6$ & $902954.5$ & $  0.4$ & $899778.5$ & $  0.0$ & $900688.8$ & $  0.1$ & $884794.4$ & $ -1.6$ \\
 $ 81$ & $\mathrm{2s^2\,2p^3\,(^2P_1)\,3d  }$ & $\mathrm{^3D_2^{o}}$ & $901473.4$ & $905155.6$ & $  0.4$ & $902031.1$ & $  0.1$ & $903539.7$ & $  0.2$ & $888375.1$ & $ -1.5$ \\
 $ 82$ & $\mathrm{2s^2\,2p^3\,(^2P_1)\,3d  }$ & $\mathrm{^3D_1^{o}}$ & $902765.0$ & $905541.2$ & $  0.3$ & $902562.6$ & $ -0.0$ & $904375.9$ & $  0.2$ & $888583.6$ & $ -1.6$ \\
 $ 83$ & $\mathrm{2s^2\,2p^3\,(^2P_1)\,3d  }$ & $\mathrm{^3D_3^{o}}$ & $902151.6$ & $905770.5$ & $  0.4$ & $902680.1$ & $  0.1$ & $904428.6$ & $  0.3$ & $888586.9$ & $ -1.5$ \\
 $ 84$ & $\mathrm{2s^2\,2p^3\,(^2P_1)\,3d  }$ & $\mathrm{^1D_2^{o}}$ & $902508.2$ & $906153.4$ & $  0.4$ & $903055.2$ & $  0.1$ & $904675.5$ & $  0.2$ & $890381.1$ & $ -1.3$ \\
 $ 85$ & $\mathrm{2s^2\,2p^3\,(^2P_1)\,3d  }$ & $\mathrm{^1F_3^{o}}$ & $905369.1$ & $909276.1$ & $  0.4$ & $905956.6$ & $  0.1$ & $907419.0$ & $  0.2$ & $892739.4$ & $ -1.4$ \\
 $ 86$ & $\mathrm{2s^2\,2p^3\,(^2P_1)\,3d  }$ & $\mathrm{^1P_1^{o}}$ & $914499.2$ & $918043.2$ & $  0.4$ & $915111.9$ & $  0.1$ & $915287.1$ & $  0.1$ & $903386.1$ & $ -1.2$ \\
\hline
\end{longtable*}

The present calculated excitation energies of the target states
are compared with the available measured values from
the NIST compilation~\cite{nist2016} and other models in \hbox{Table~\ref{tab:energies}}.
In addition to the \hbox{BSR-86 (TS)} and GRASP-86 structure results, we include
the extensive MCHF calculations by Froese-Fischer and Tachiev~\cite{froese-fischer2004}.
The latter calculations were carried out with an extremely large set
of configurations and a careful analysis of the convergence was performed.
It is generally accepted that this work represents the most accurate calculation
for the structure of the lowest excited levels of several \hbox{O-like} ions, including $\mathrm{Mg}^{4+}$.
As seen from the table, the MCHF and \hbox{BSR-86 (TS)} calculations show the best agreement with the NIST-recommended data,
whereas the DARC-86 energies lack considerably in accuracy due to
the limited CI expansions used in the target wave\-functions.
This indicates large correlation corrections due to CI effects in the case of interest.
Except for the fine-structure splitting of the $\mathrm{2s^2\,2p^4}$ ground-state configuration,
the present \hbox{BSR-86} excitation energies are also in close agreement with experiment,
with deviations of generally less than~$0.5\%$.
The larger difference in the fine-structure splitting for the lowest $\mathrm{2s^2\,2p^4}$ configuration is most likely
due to restricting the configuration expansion as described above.
In spite of the slightly larger deviations in comparison to the MCHF results,
we demonstrate below that our wave\-functions accurately represent the main correlation corrections,
as well as the interaction between different Rydberg series and term-dependence
effects.

\begin{figure}
   \includegraphics[width=1.00\columnwidth,clip]{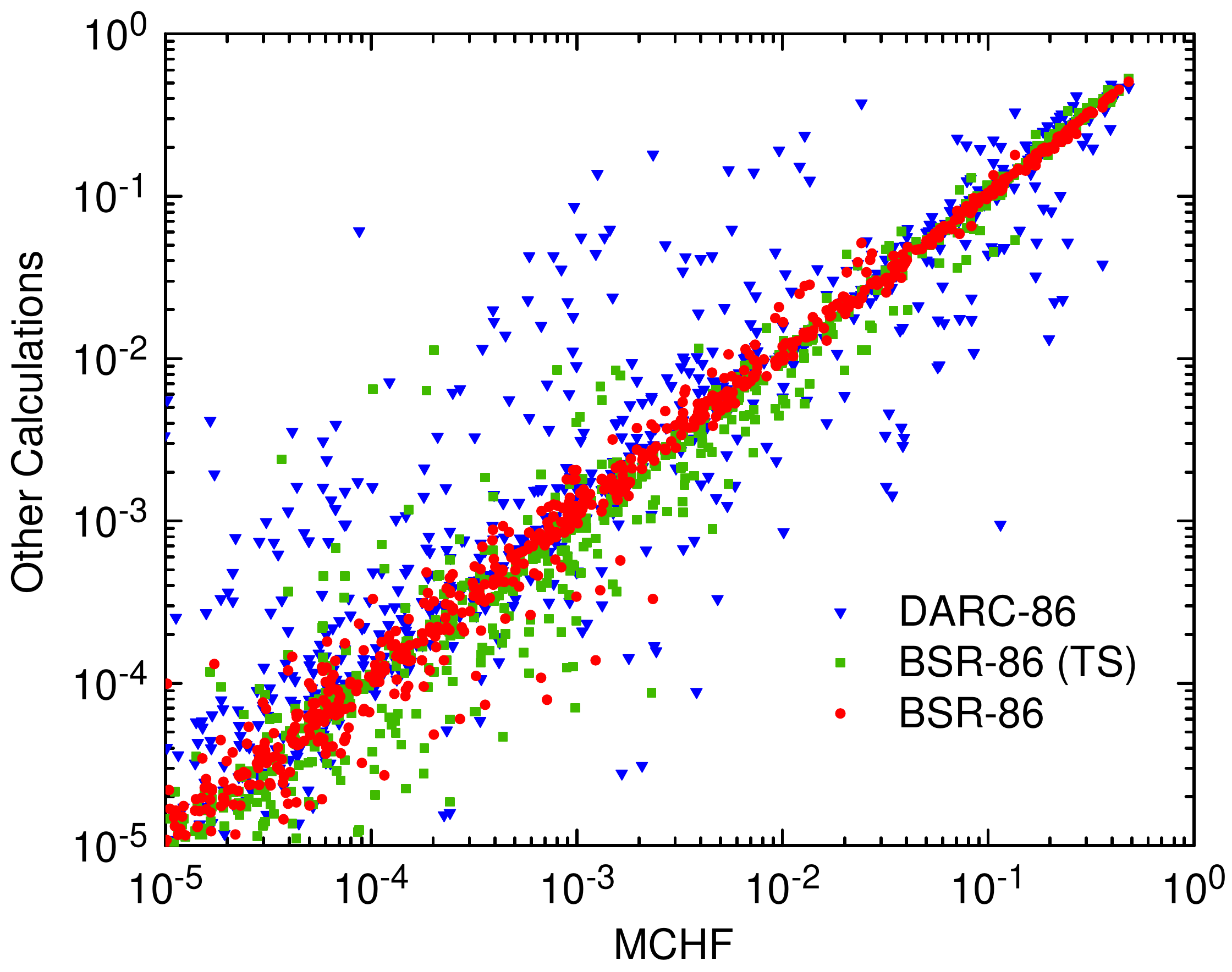}
   \caption{\label{fig:fvalues} Comparison of $f$-values (dimensionless) from the structure 
     models used in the present \hbox{BSR},
     \hbox{BSR-86 (TS)}, and \hbox{DARC-86} calculations
     with the MCHF results~\cite{froese-fischer2004}.
     $x$-axis: $f$ results for a certain transition calculated with the MCHF
     method \cite{froese-fischer2004};
     $y$-axis: $gf$ results for the same transition calculated with 
     $\bullet$ BSR (present work),
     $\blacksquare$ \hbox{BSR-86 (TS)}~\cite{tayal2015} and
     $\blacktriangledown$ \hbox{DARC-86}~\cite{aggarwal2016b}.
     }
\end{figure}

Figure~\ref{fig:fvalues} exhibits a comparison of the MCHF oscillator strengths
with those from the structure calculations used in the present \hbox{BSR} and
the previous \hbox{BSR-86 (TS)} models~\cite{tayal2015}, as well as
the DARC-86 calculations~\cite{aggarwal2016b} for transitions between all
86 target levels presented in Table~\ref{tab:energies}.
In the absence of experimental data, we consider the MCHF calculations~\cite{froese-fischer2004} 
as the reference to be compared with.
These data are also the recommended values in the NIST compilation~\cite{nist2016}
as the most accurate currently available structure data for the $\mathrm{Mg}^{4+}$ ion.
There is very good agreement between the present results and the MCHF predictions for all transitions,
including the very weak ones with small \hbox{$f$-values}.
This suggests similarly accurate configuration mixing in both calculations.
The \hbox{$f$-values} used in the \hbox{BSR-86 (TS)} model show somewhat larger deviations, mainly
for the weak transitions.
The GRASP results for the \hbox{DARC-86} model, on the other hand, differ considerably from the MCHF
values for many transitions,
including even relatively strong ones with \hbox{$f$-values} larger than $10^{-3}$.
This is a clear indication that the GRASP target wave\-functions miss important CI corrections.

\begin{figure}
   \includegraphics[width=1.00\columnwidth,clip]{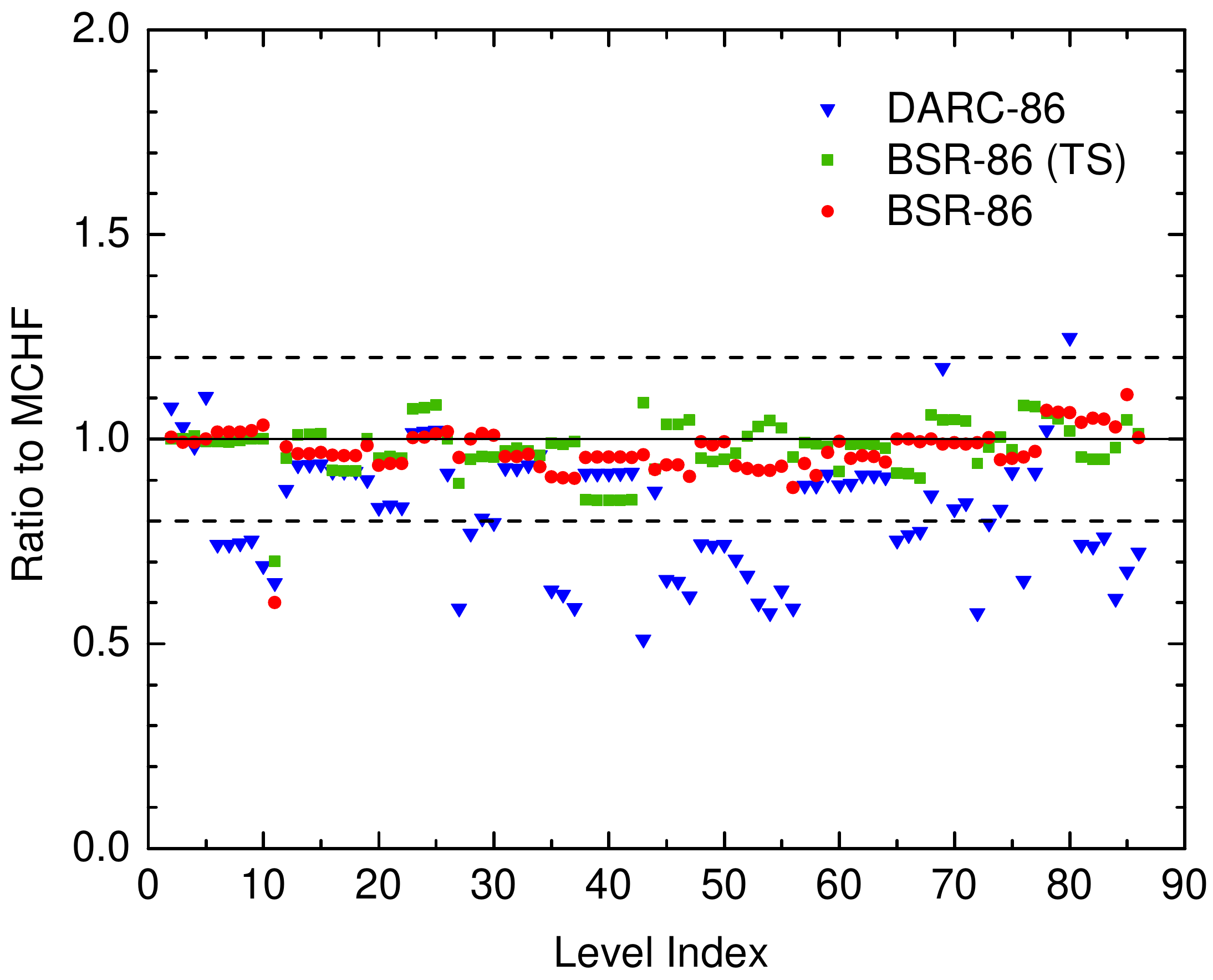}
   \caption{\label{fig:tau} Lifetimes from the structure models used in the present \hbox{BSR}, \hbox{BSR-86 (TS)},
            and \hbox{DARC-86}
            calculations, normalized to the MCHF predictions~\cite{froese-fischer2004}.
            The dashed lines indicate the bands of $\pm 20\%$ deviation. }
\end{figure}

Variations in the $f$-values also lead to different predictions of the lifetimes.
A comparison of the lifetimes from the various calculations is shown in Fig.~\ref{fig:tau}.
As expected, the present BSR lifetimes are in closest agreement with the MCHF results~\cite{froese-fischer2004}.
Except for the weak transition ($1-11$) to the $\mathrm{(2s^2\,2p^3(^4S)\,3s)^5S_2^{\rm o}}$ state,
all our values agree with the MCHF results within $10\%$.
The lifetimes of \hbox{BSR-86 (TS)}~\cite{tayal2015} agree within $20\%$,
whereas the GRASP results of DARC-86~\cite{aggarwal2016b} again exhibit much larger deviations for many levels,
both for low-lying and high-lying states.
This indicates, once again, that the \hbox{GRASP} configuration expansions are far from
converged.  Consequently, the accuracy of the predicted collision strengths from the \hbox{DARC-86} model with
these inferior structure data as input are highly questionable.
This will be further discussed below.

\subsection{\label{subsec:coll} Scattering calculations} 

We performed two sets of scattering calculations.
In the first model, \hbox{BSR-86}, we selected the same set of target states as in the previous \hbox{BSR-86 (TS)}
and \hbox{DARC-86}  calculations.
Comparison between the predictions from these three \hbox{86-state} models
allows us to directly draw conclusions regarding the sensitivity
of the final rate coefficients to the target structure description.
In our second model we additionally included the states from the $\mathrm{2s^2\,2p^4\,4}l$ and
$\mathrm{2s\,2p^5\,3}l$ configurations, yielding 316 levels overall.
This \hbox{BSR-316} model
allows us to check the convergence of the CC expansion,
especially for transitions to higher-lying states.

The close-coupling equations were solved by means of the \hbox{R-matrix} method, using a
parallelized version of the BSR complex \cite{zatsarinny2006}.
The distinctive feature of the method is the use of \hbox{B-splines} as a universal basis set
to represent the scattering orbitals in the inner region of $r \le a$.
Hence, the \hbox{R-matrix expansion} in this region takes the form
\begin{eqnarray}\label{eq:RM}
& & \Psi_k(x_1,\ldots,x_{N+1})  =  \nonumber\\
& & {\cal A} \sum_{ij} \bar{\Phi}_i(x_1,\ldots,x_N;\hat{\mathbf{r}}_{N+1}\sigma_{N+1})
\,r_{N+1}^{-1}\,B_j(r_{N+1})\,a_{ijk} \nonumber\\
& & +  \sum_i \chi_i(x_1,\ldots,x_{N+1})\,b_{ik}\,.
\end{eqnarray}
Here ${\cal A}$ denotes
the anti\-symmetrization operator, the $\bar{\Phi}_{i}$ are the channel functions constructed from the
$N$-electron target states and the angular and spin coordinates of the projectile, and
the splines $B_{j}(r)$ represent the radial part of the continuum orbitals.
The $\chi_i$ are additional $(N\!+\!1)$-electron bound states.
In standard \hbox{$R$-matrix} calculations \cite{burke2011}, the latter are included
one configuration at a time to ensure completeness of the expansion
when compensating for orthogonality constraints imposed on
the continuum orbitals.
The use of non\-orthogonal one-electron radial functions in the BSR method, on the other hand,
allows us to avoid these configurations for compensating orthogonality restrictions,
thereby avoiding the pseudo\-resonance structure that may appear
in calculations with an extensive number of bound channels $\chi_i$ in the CC expansion.

In the inner region, we used the same \hbox{B-spline} set as for the target description
described above. The maximum interval in the \hbox{B-spline} grid was $0.25\,a_0$.
This is sufficient to cover electron scattering energies up to $50\,\mathrm{Ry}$.
Numerical calculations were performed for 60 partial waves, with
total electronic angular momentum $J = 1/2 - 59/2$, for both even and odd parities.
In the \hbox{BSR-86} model, the maximum number of channels in a single partial wave was $388$
while it was $1568$ in \hbox{BSR-316}.  With a basis size of 82 \hbox{B-splines},
this required the diagonalization of matrices with dimensions up to about $32\,000$ and
$130\,000$, respectively.
The former calculations could still be performed on modern desk\-top machines, while the latter
were carried out with parallelized versions of the BSR complex,
using super\-computers with distributed memory.

For the outer region we employed the parallel version of the PSTGF
program (\url{http://www.apap-network.org/codes.html}).
In the resonance region for impact energies below the excitation energy of the highest level
included in the CC expansion, we used a fine energy step
of $10^{-5}\,z^2\,\mathrm{Ry}$, with $z=4$ as the target ionic charge, to
properly map those resonances.
For energies above the highest excitation threshold
included in the CC expansion, the collision strengths vary
smoothly, and hence we chose a coarser step of $10^{-2}\,z^2\,\mathrm{Ry}$.
Altogether, $51416$ energies for the colliding electron were considered in the \hbox{BSR-86} model
and $71376$ energies in the \hbox{BSR-316}  model.
We calculated collision strengths up to~$50\,\mathrm{Ry}$, which is about five times the ionization
potential of $\mathrm{Mg}^{4+}$.
For even higher energies, if needed, we extrapolated $\Omega$ using the well-known asymptotic 
energy dependence of the various transitions.
To obtain effective collision strengths~$\Upsilon(T_\mathrm{e})$, we convoluted
$\Omega$ with a Maxwellian distribution for an electron temperature $T_\mathrm{e}$, i.e.,
\begin{equation}
   \Upsilon_{i-j}(T_\mathrm{e})\ =\ \int_{E_{th}}^{\infty} {\rm d} E\,\Omega_{i-j}(E)\,
   \exp \left( \frac{E-E_{th}}{kT_\mathrm{e}} \right) \,.
\end{equation}
Here $E_{th}$ is the $i-j$ transition energy and $k$ is the Boltz\-mann constant.
We calculated $\Upsilon$ for temperatures between $10^{3}$ and $10^{6}\,\mathrm{K}$.
The entire table with effective collision strengths for all temperatures and
transitions included in the \hbox{BSR-316} model can be found online in the Supplemental
Material provided with the present manuscript.

\section{\label{sec:results} Results and Discussion}

\subsection{BSR-86 calculations}

\begin{figure}[b!]
\center
   \includegraphics[width=0.80\columnwidth,clip]{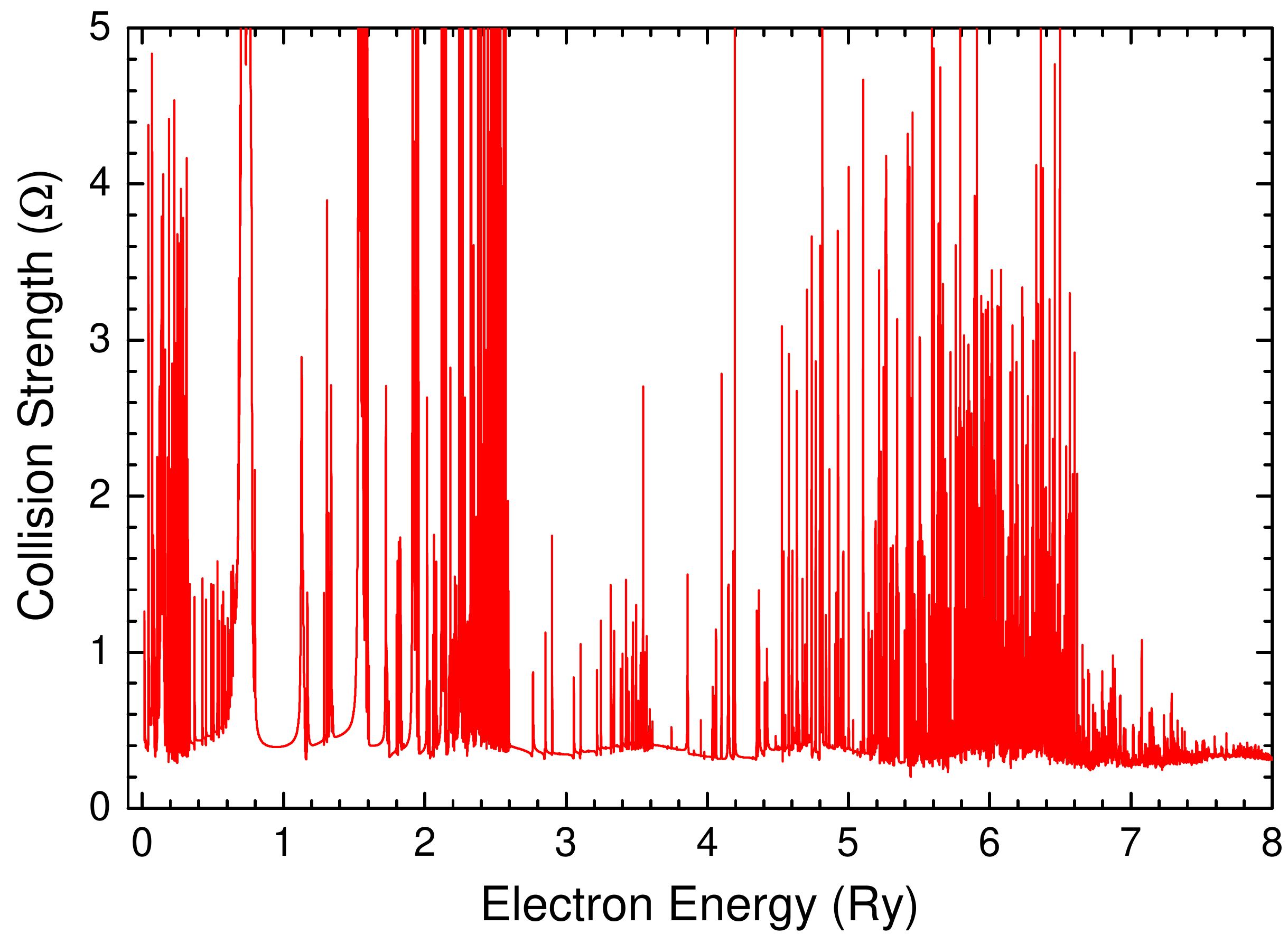}
   \includegraphics[width=0.80\columnwidth,clip]{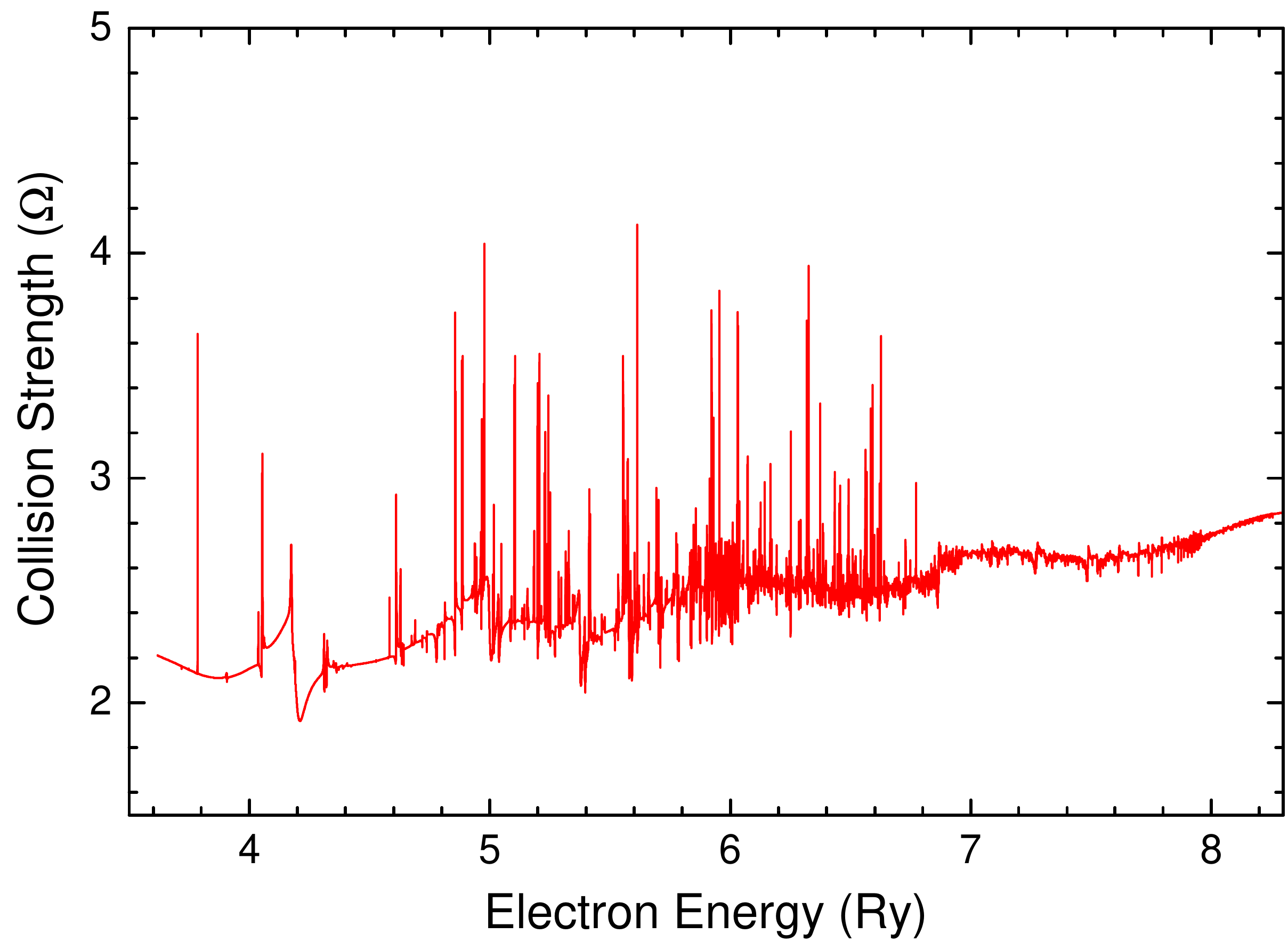}
   \caption{\label{fig:TR12+TR49} Collision strength $\Omega$ (dimensionless) for electron-impact
      excitation of the \hbox{$1-2$} \hbox{$\mathrm{(2s^2\,2p^4\,^3P_2 - 2s^2\,2p^4\,^3P_1)}$}
      (top) and \hbox{$4-9$} \hbox{$\mathrm{(2s^2\,2p^4\,^1D_2 - 2s\,2p^5\,^1P_1^{\mathrm{o}})}$} (bottom)
      transitions in $\mathrm{Mg}^{4+}$, as obtained in the present \hbox{BSR-86} model.}
\end{figure}

\begin{figure*}[t!]
\centering
      \includegraphics[width=0.32\textwidth,clip]{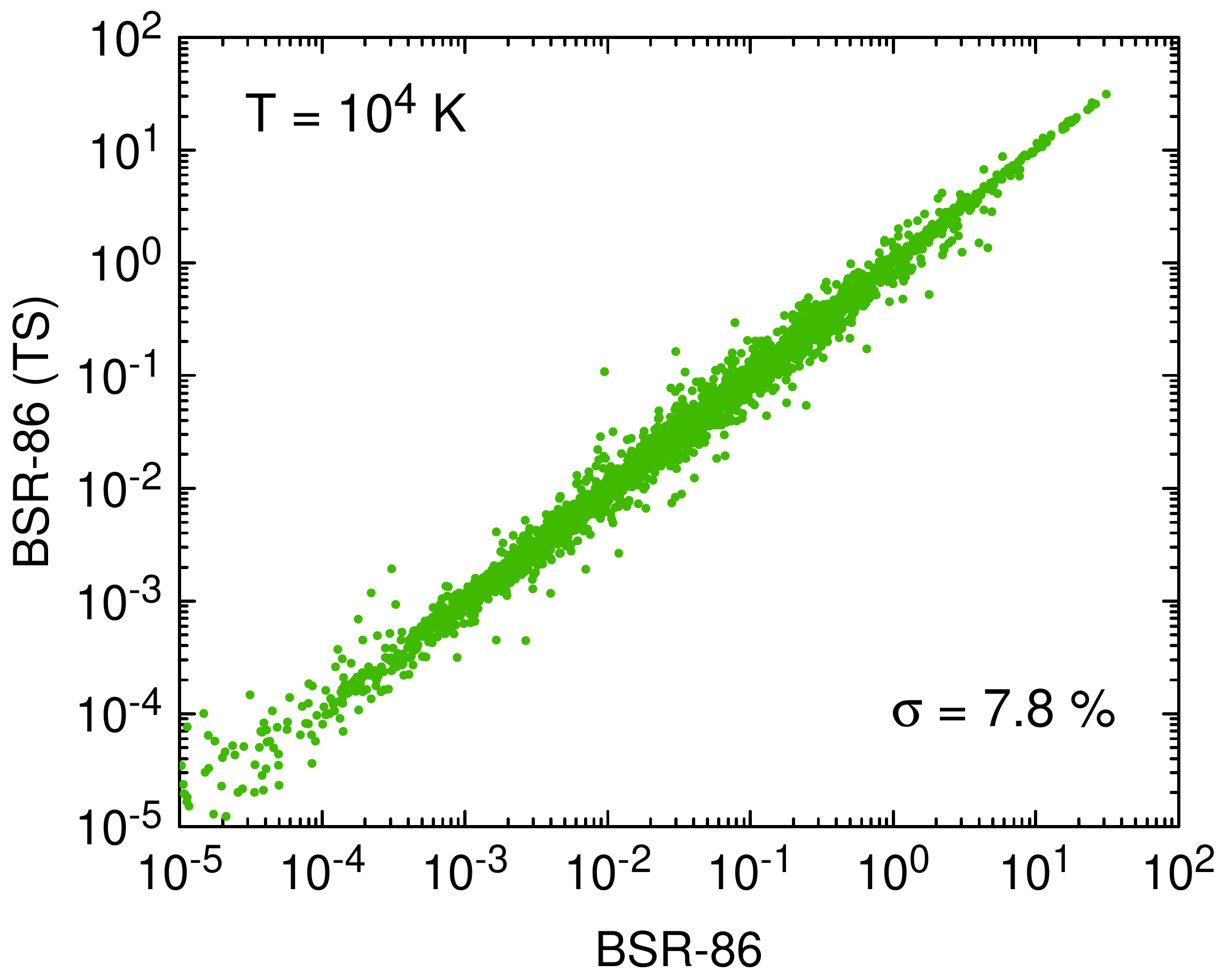}
      \includegraphics[width=0.32\textwidth,clip]{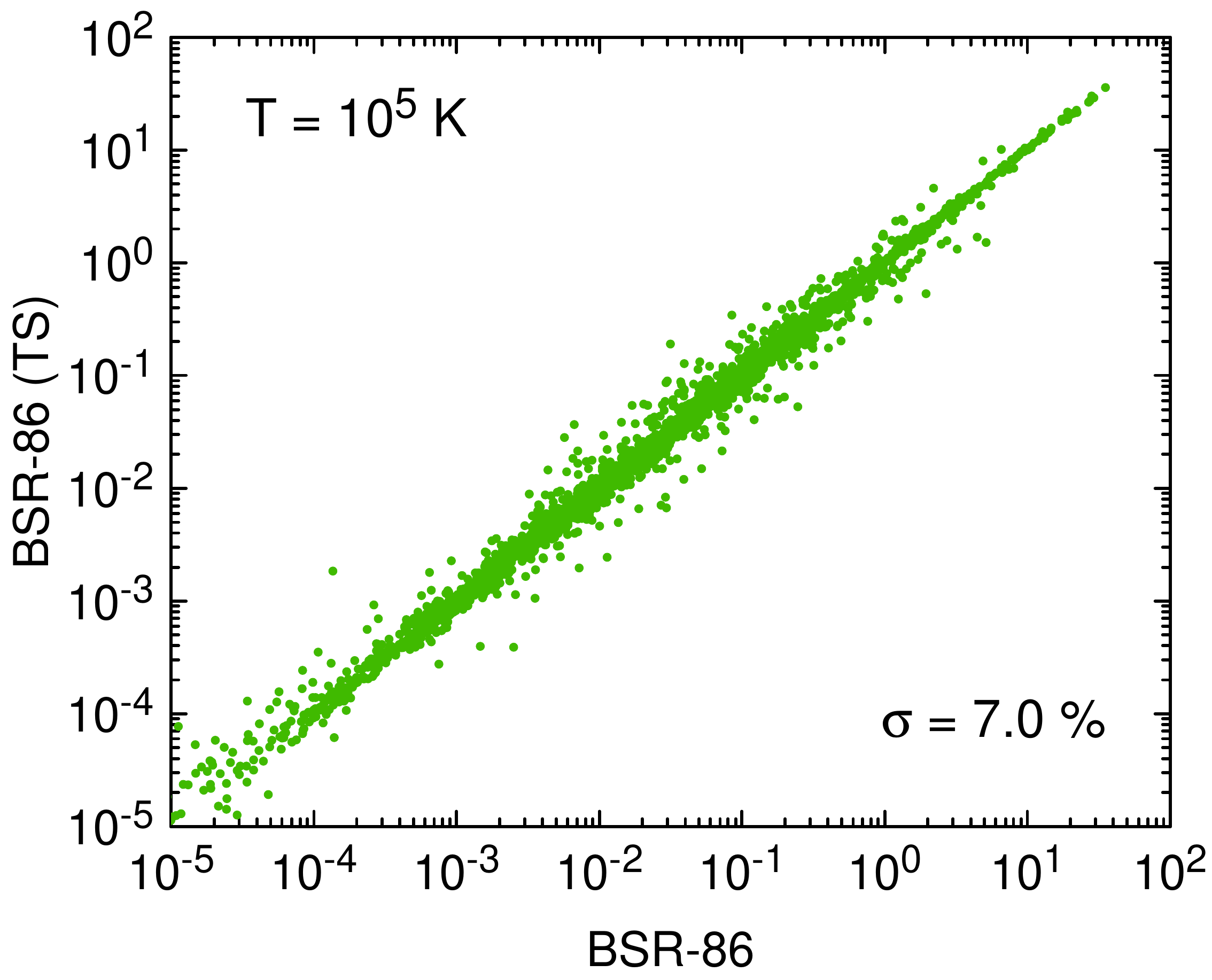}
      \includegraphics[width=0.32\textwidth,clip]{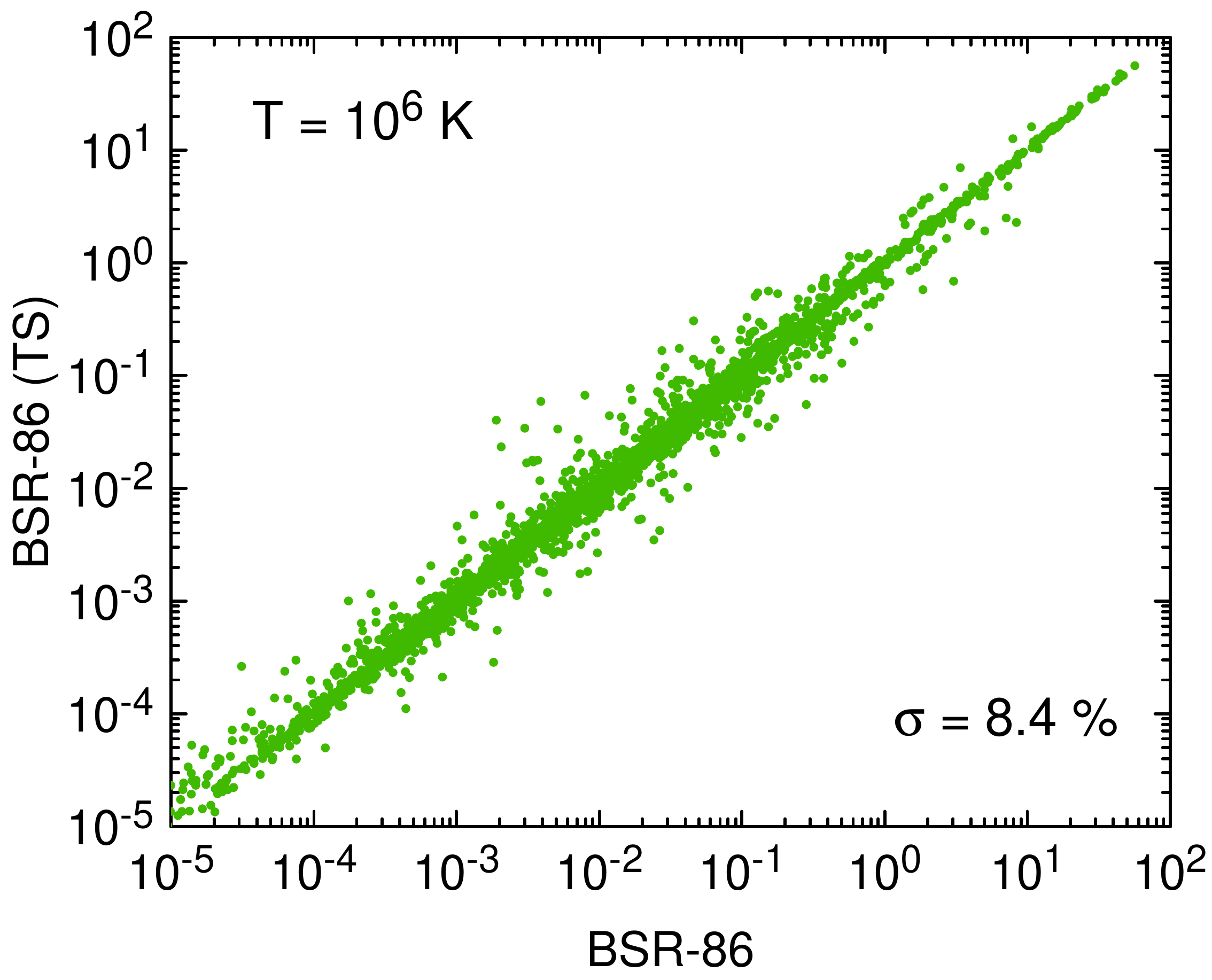}
      \includegraphics[width=0.32\textwidth,clip]{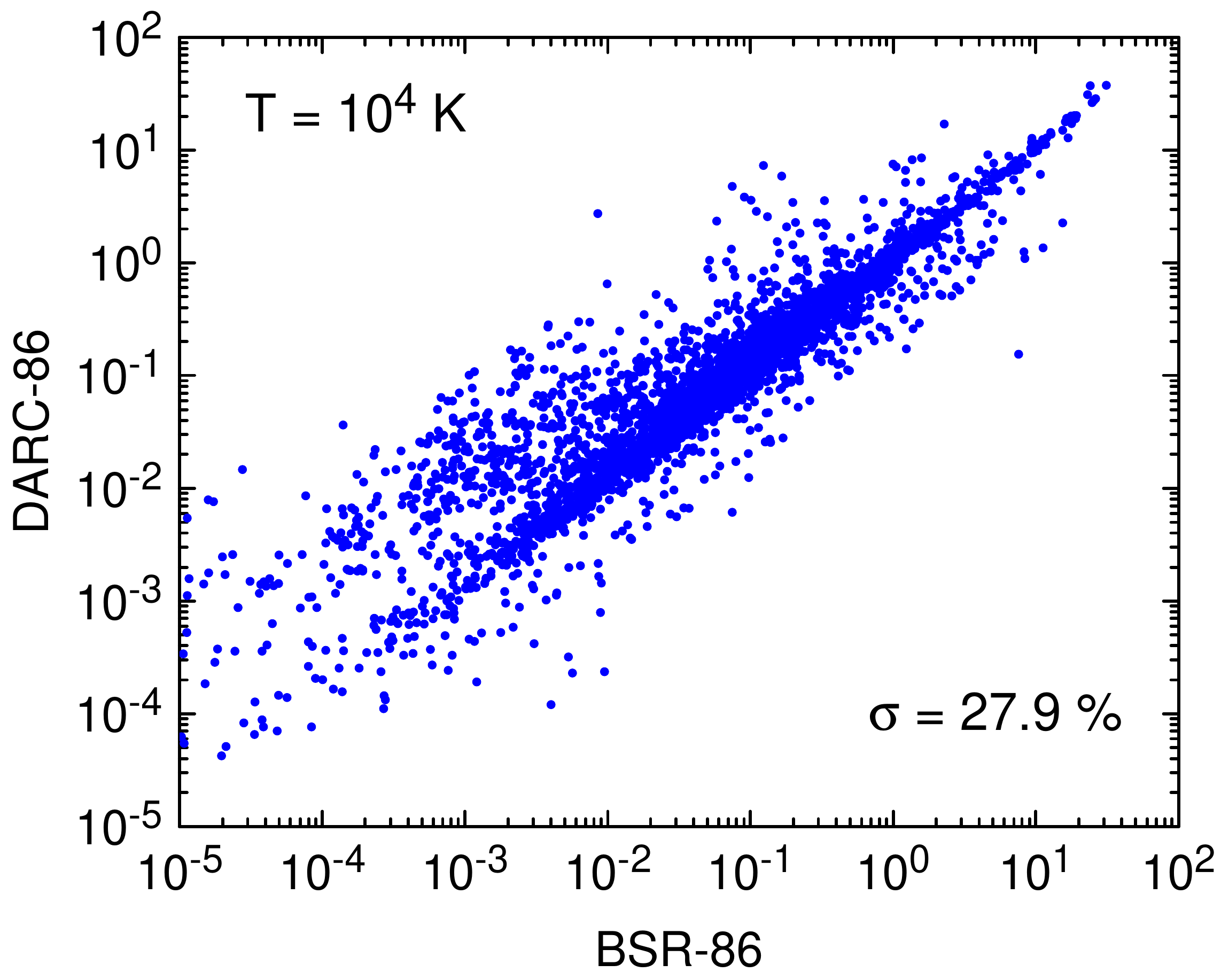}
      \includegraphics[width=0.32\textwidth,clip]{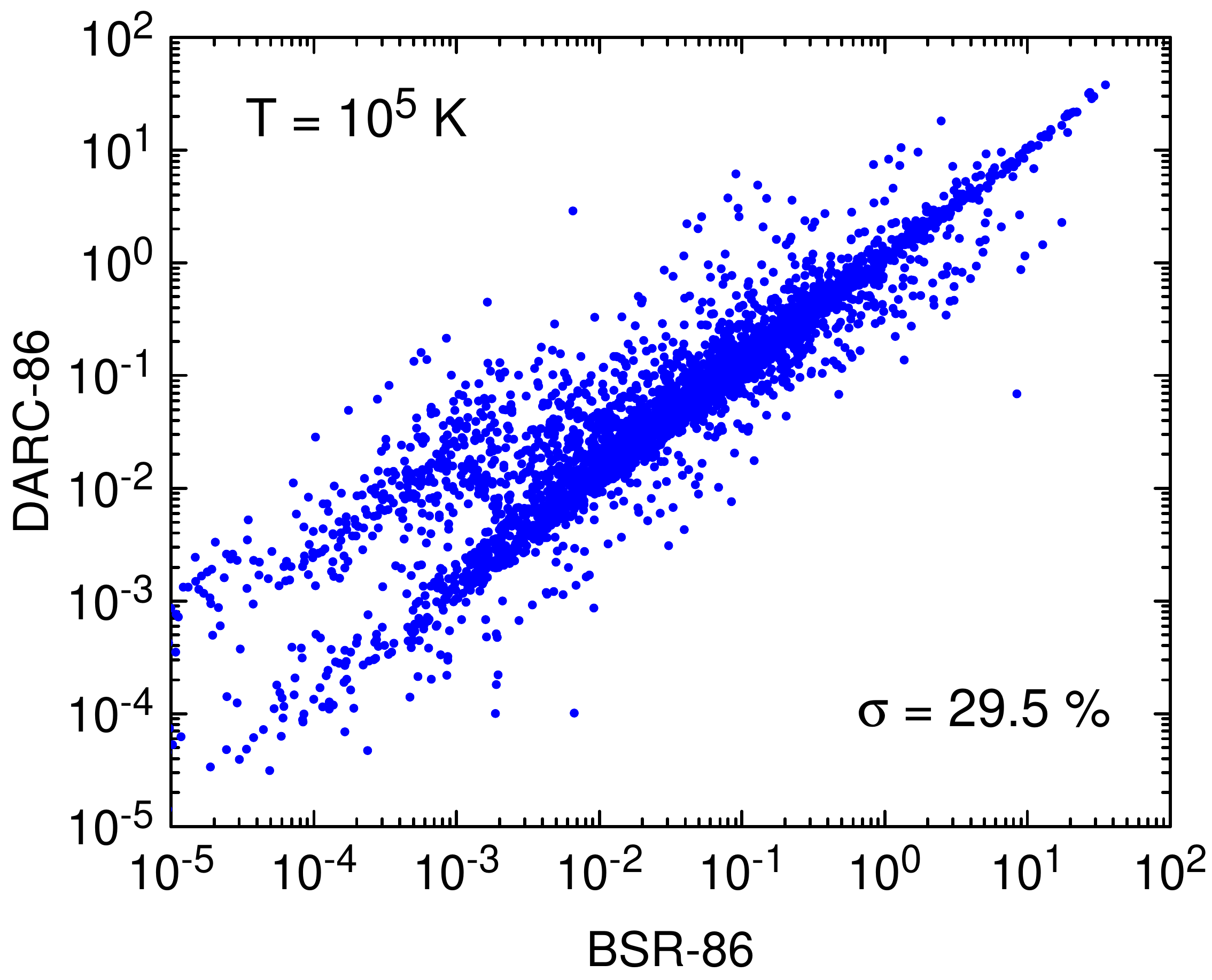}
      \includegraphics[width=0.32\textwidth,clip]{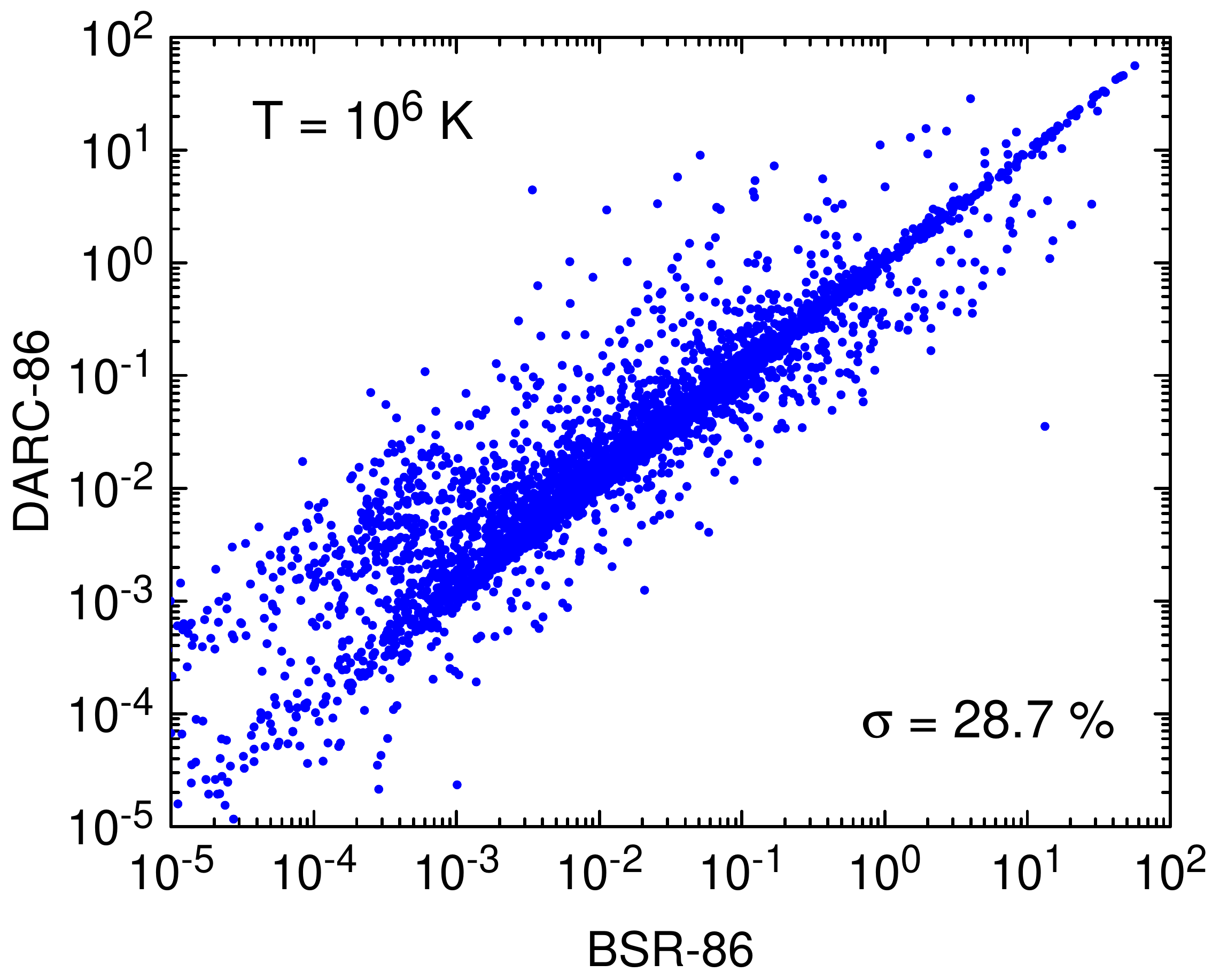}
      \includegraphics[width=0.32\textwidth,clip]{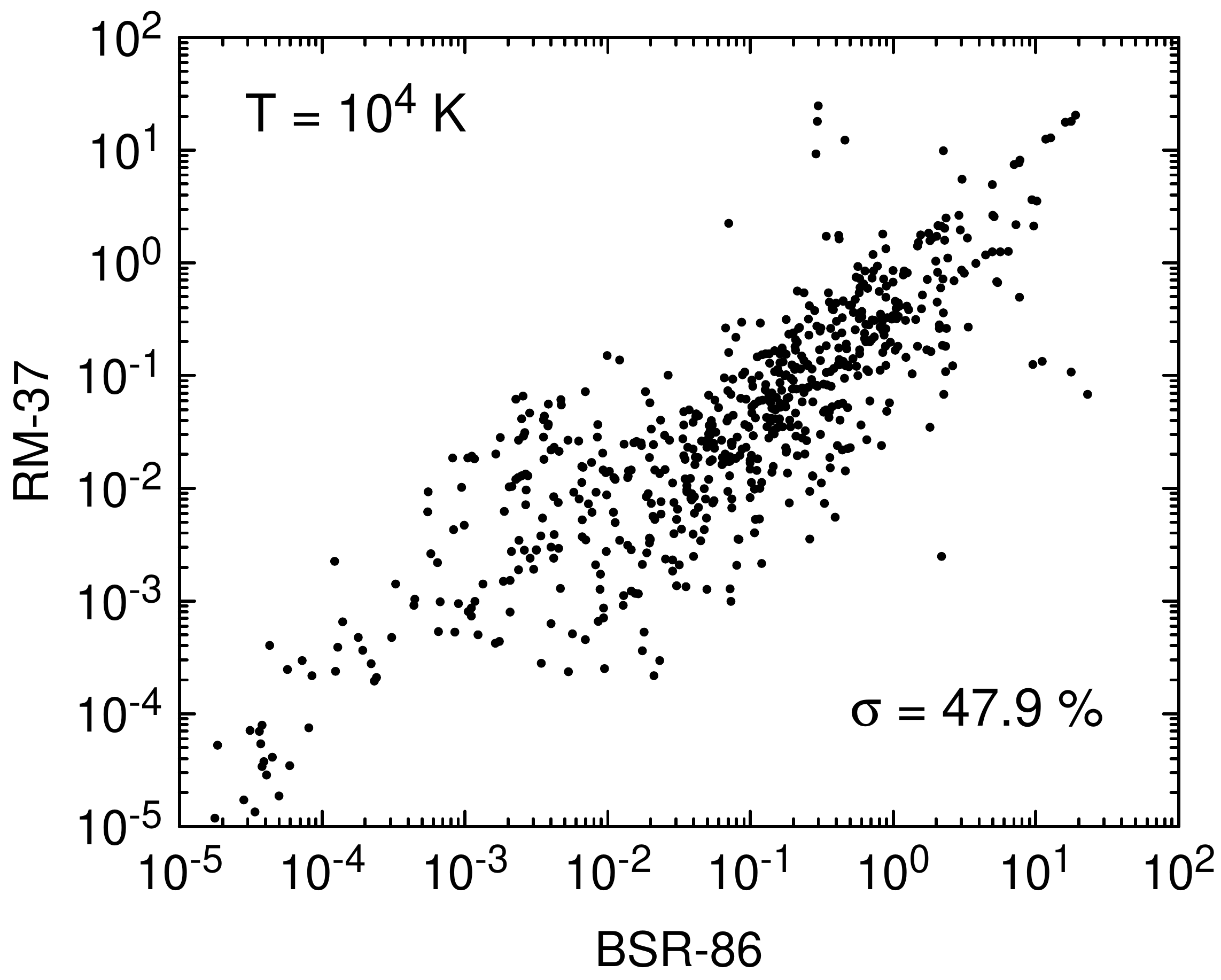}
      \includegraphics[width=0.32\textwidth,clip]{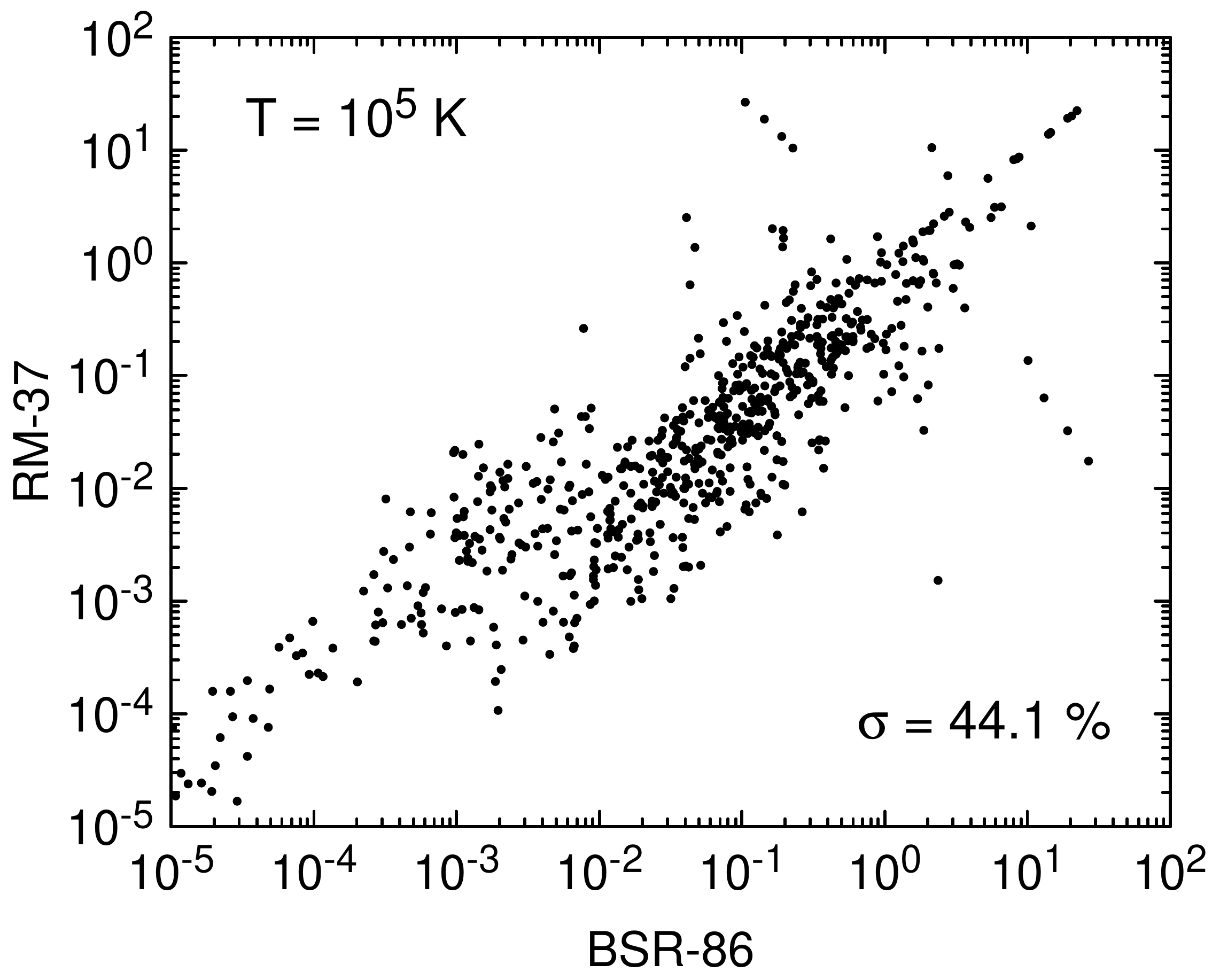}
      \includegraphics[width=0.32\textwidth,clip]{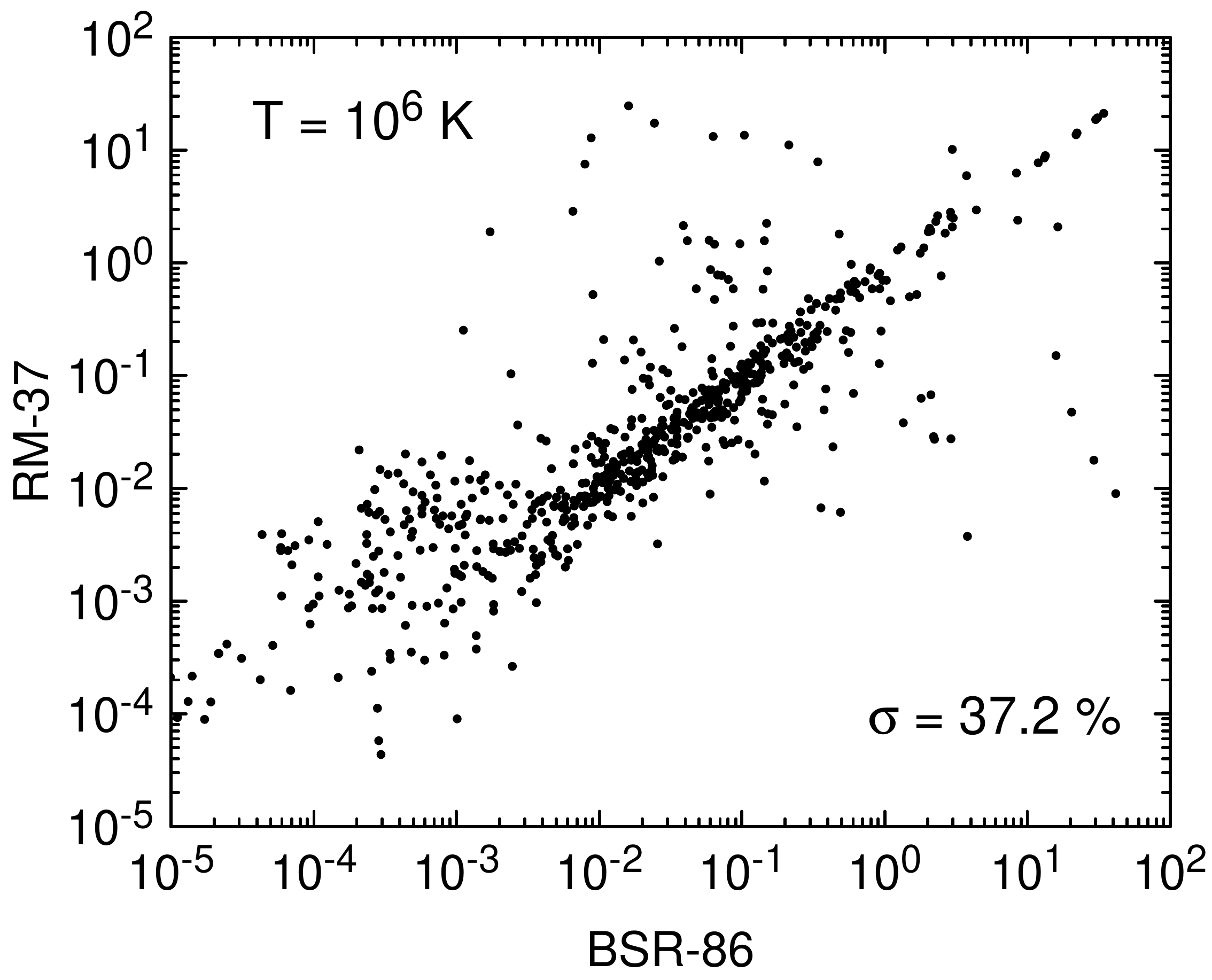} 
\caption{Comparison of effective collision strengths $\Upsilon$ (dimensionless) 
         obtained in the present \hbox{BSR-86} model with
         \hbox{BSR-86 (TS)} (upper panels), \hbox{DARC-86} (middle panels), and \hbox{RM-37} (lower panels) for
         three temperatures. 
         Also indicated in the panels is the average deviation~$\sigma$ from the \hbox{BSR-86} results.
         $x$-axis: $\Upsilon$ results for a certain transition calculated with present BSR-86 model;
         $y$-axis: $\Upsilon$ results for the same transition calculated with other previous model.
         }
\label{fig:COMP}
\end{figure*}

As our first example, Fig.~\ref{fig:TR12+TR49} exhibits the resonance structure of the
collision strength~$\Omega$ for the forbidden M1 transition
$\mathrm{(2s^2\,2p^4\,^3P_2 - 2s^2\,2p^4\,^3P_1)}$
and for the dipole-allowed transition
$\mathrm{(2s^2\,2p^4\,^1D_2 - 2s\,2p^5\,^1P_1^{\mathrm{o}})}$.
We see the typical resonance structure for electron$-$ion scattering that
consists of numerous closed-channel resonances, most of which are narrow.
This resonance structure provides the dominant part of the collision strength
for weak forbidden transitions, but it only yields a relatively small contribution to strong
dipole-allowed transitions.
Visual comparison with similar plots from the \hbox{DARC-86}  calculation~\cite{aggarwal2016b} shows good
qualitative agreement between the calculations for these two cases.
This also indicates that both calculations sufficiently resolve
the rich structure.  Hence,
further increasing the number of energy points would not lead to noticeable corrections.

A global comparison between the present \hbox{BSR-86} results and the effective collision strengths obtained
previously is presented in Fig.~\ref{fig:COMP} at three different temperatures.
The best agreement is observed with the  \hbox{BSR-86 (TS)} predictions~\cite{tayal2015}.
The average relative deviation is around $8\%$, which is considered a very good agreement
for such type of calculation.
Much worse agreement (with average deviations of almost $30\%$) is observed with the \hbox{DARC-86}
calculations~\cite{aggarwal2016b}.
To some extent the dispersions at all three temperatures are similar to the ones seen for the $f$-values
in Fig.~\ref{fig:fvalues}. This is a clear indication that the target structure description is the
principal source for the differences.
For completeness, Fig.~\ref{fig:COMP} also contains a comparison with the earlier \hbox{RM-37}  
results~\cite{hudson2009b}.
In addition to only coupling the lowest 37 levels of $\mathrm{Mg}^{4+}$, the energy mesh
used in the \hbox{RM-37} calculation is much coarser than in the other calculations.
This may lead to a poor account of the resonance contributions.
As discussed below, some of the RM-37 collision strengths exhibit apparent pseudo\-resonance structure.    
As a result, the average deviation of $44\%$ from the present results is understandable.

As shown by the above comparison, our present work confirms the validity of previous 
\hbox{BSR-86 (TS)} results~\cite{tayal2015}.
The differences are reasonable and can be explained by the different structure descriptions.
This conclusion contradicts the statements of Aggarwal and Keenan~\cite{aggarwal2016b},
who declared that the \hbox{BSR-86 (TS)} results are incorrect and assessed the
corresponding $\Upsilon$-values as unreliable.
They also expressed doubts about the ability of the BSR approach to avoid
pseudo\-resonances, and thereby to generate accurate results in general.

The principal argument of Aggarwal and Keenan~\cite{aggarwal2016b} is that the BSR $\Upsilon$-values show a
big ``hump'' at $10^5\,\mathrm{K}$ for almost all transitions to the state \#86,
and that the collision strengths appear to be very much underestimated at low(er) temperatures.
Such a dependence of $\Upsilon$ looks un\-physical and hence the
appearance of pseudo\-resonances in the \hbox{BSR-86 (TS)} calculation was suggested.
Since the suspicion of Aggarwal and Keenan regarding the correctness of the results for
state \#86 appears to be warranted, we decided to perform a detailed analysis
and requested the original values for the collision strength $\Omega$ from the authors of~\cite{tayal2015}.
We found that the dense energy mesh in the  \hbox{BSR-86 (TS)}
calculations ends right below the threshold of this last state and that there is a relatively large gap
of $0.02\,\mathrm{Ry}$ between the excitation threshold and the next energy point.
As a result, the corresponding near-threshold portion of the excitation function is not accounted
for properly when calculating $\Upsilon$ for this state.
This finding explains the observations of Aggarwal and Keenan,
but the problem is actually {\em limited} to transitions to state~\#86.
A similar analysis for other states of $\mathrm{Mg}^{4+}$
does not confirm the conclusions of Aggarwal and Keenan.
The extrapolation of their conclusions from a single level to all others is simply not warranted.

\begin{figure}
\center
   \includegraphics[width=0.80\columnwidth,clip]{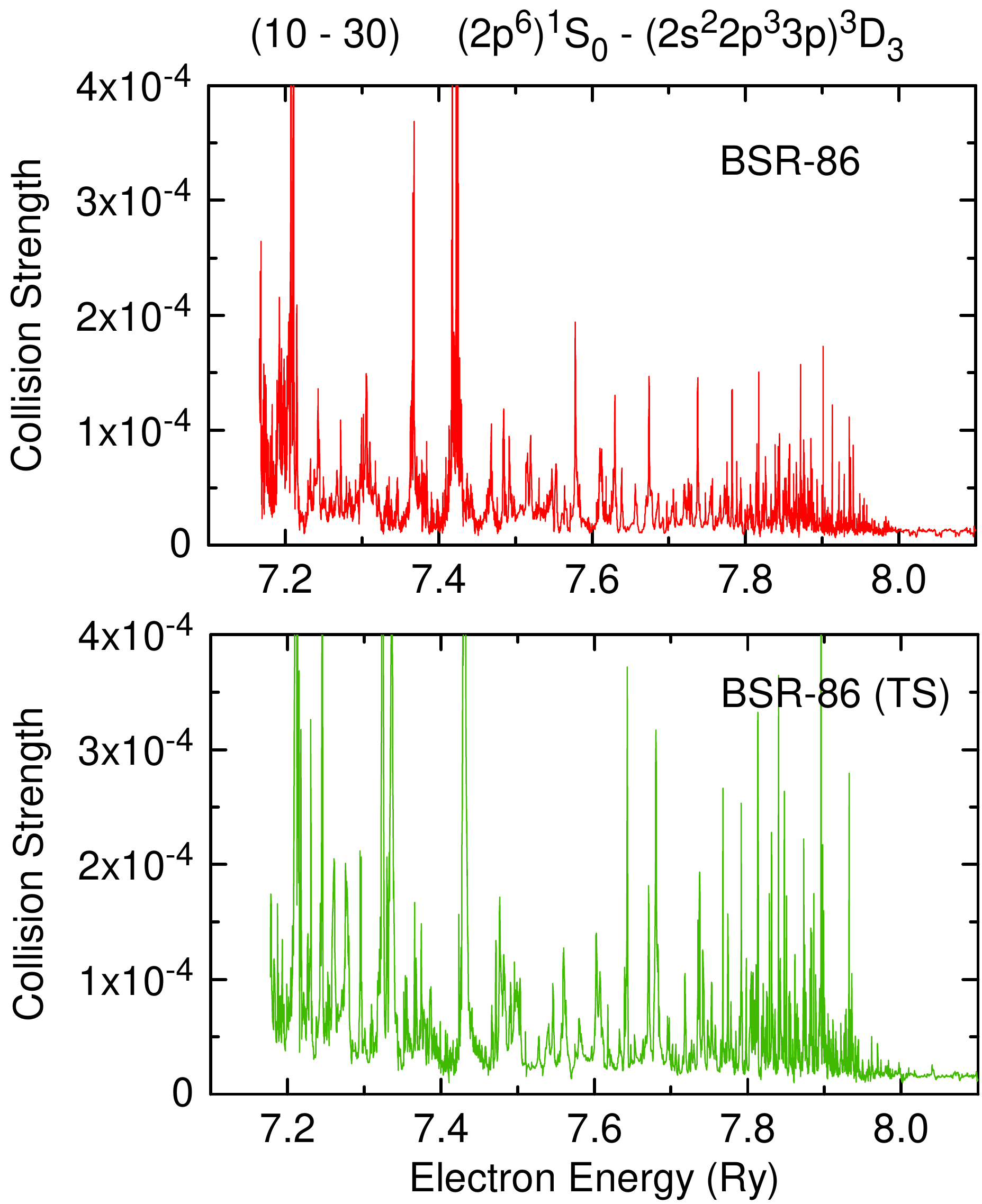}
   \caption{\label{fig:TR1030} Collision strength $\Omega$ for electron-impact
      excitation of the $10-30$ transition in the \hbox{BSR-86} (top) and \hbox{BSR-86 (TS)} (bottom) models.}
\end{figure}

Also,
the suggested appearance of pseudo\-resonances in the BSR calculations contradicts the
findings of Aggarwal and Keenan~\cite{aggarwal2016b} that for many transitions (especially from levels below \#12) the
DARC $\Upsilon$-values are still much larger, up to two orders of magnitude.
The reason given is the large difference in the background collision strengths.
One such example is the \hbox{$10-30$} \hbox{($\mathrm{2p^6\,^1S_0 - 2s^2\,2p^3\,3p\,^3D_3}$)} transition,
for which Aggarwal and Keenan exhibit the collision strength in Fig.~6 of~\cite{aggarwal2016b}.
The BSR collision strength for this transition is shown in Fig.~\ref{fig:TR1030}.
We see close agreement between the \hbox{BSR-86} and  \hbox{BSR-86 (TS)}  results.
Visual comparison with the \hbox{DARC-86} predictions reveals drastic deviations in both shape and magnitude.
Whereas the BSR calculations exhibit the typical resonance structure consisting
of dense but narrow resonances, the DARC cross section is dominated  by
three intense and broad maxima, which are typical for pseudo\-resonances.
In other words, we arrive at the opposite conclusion compared to Aggarwal and Keenan
regarding the possible effect of pseudo\-resonances on the results.
It is also worth noting that the $10-30$ transition is a strongly forbidden 
three-electron-jump M3 transition whose strength depends strongly on the correlation
corrections in the underlying target description as well as close-coupling effects.

\subsection{BSR-316  calculations}

\begin{figure}
   \includegraphics[width=0.95\columnwidth,clip]{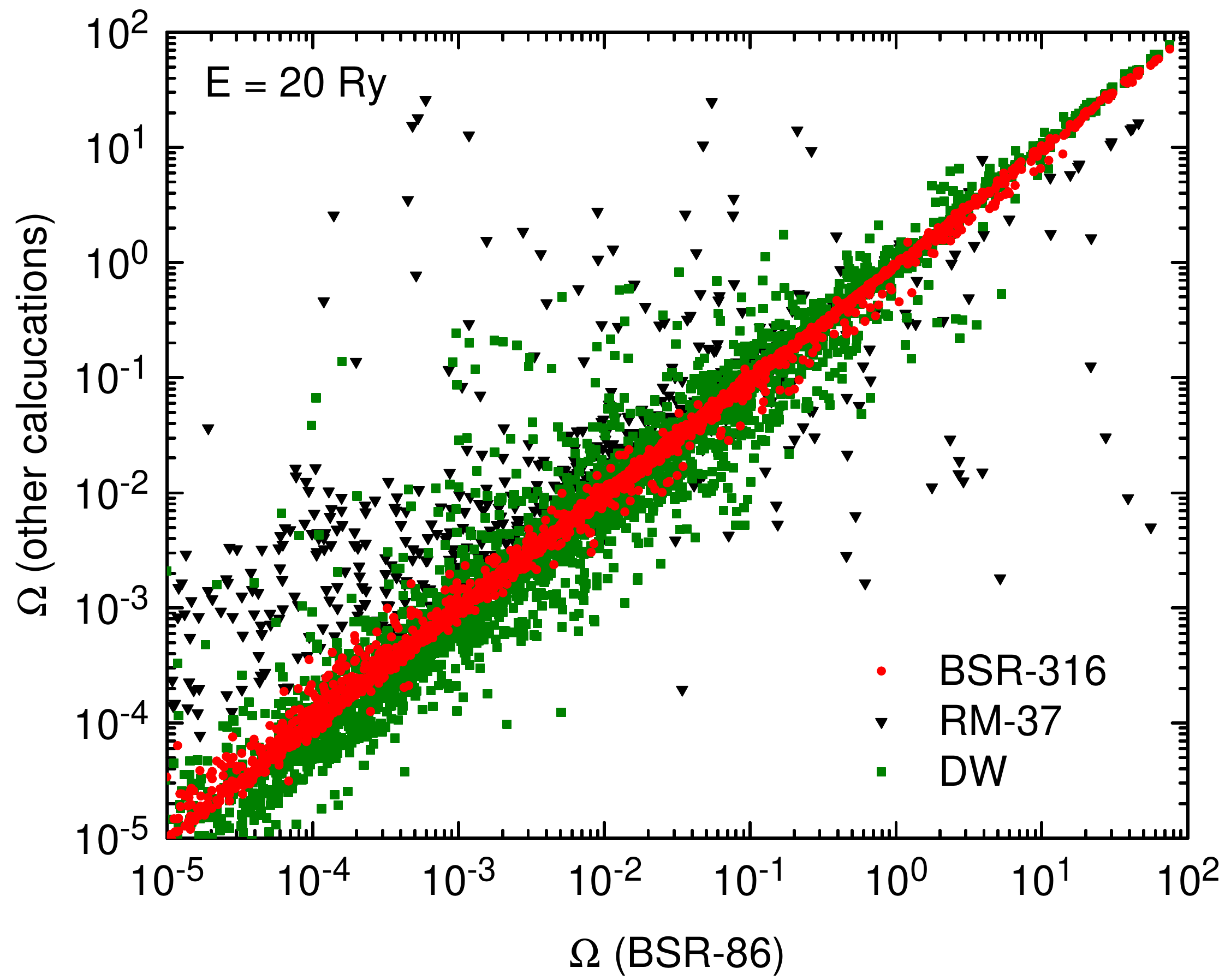}
   \caption{\label{fig:omega} Comparison of collision strengths for the \hbox{BSR-86} model with
      the DW, \hbox{RM-37}, and \hbox{BSR-316}  calculations at the electron
      energy $E=20\,\mathrm{Ry}$.
      $x$-axis: $\Omega$ results for a certain transition calculated with present \hbox{BSR-86} model;
      $y$-axis: $\Omega$ results for the same transition calculated with 
      $\bullet$ present \hbox{BSR-316} model,
      $\blacktriangledown$ \hbox{RM-37}~\cite{hudson2009b},
      $\blacksquare$ DW~\cite{bhatia2006}.
      }
\end{figure}

\begin{figure*}[t]
\centering
\includegraphics[width=0.33\textwidth,clip]{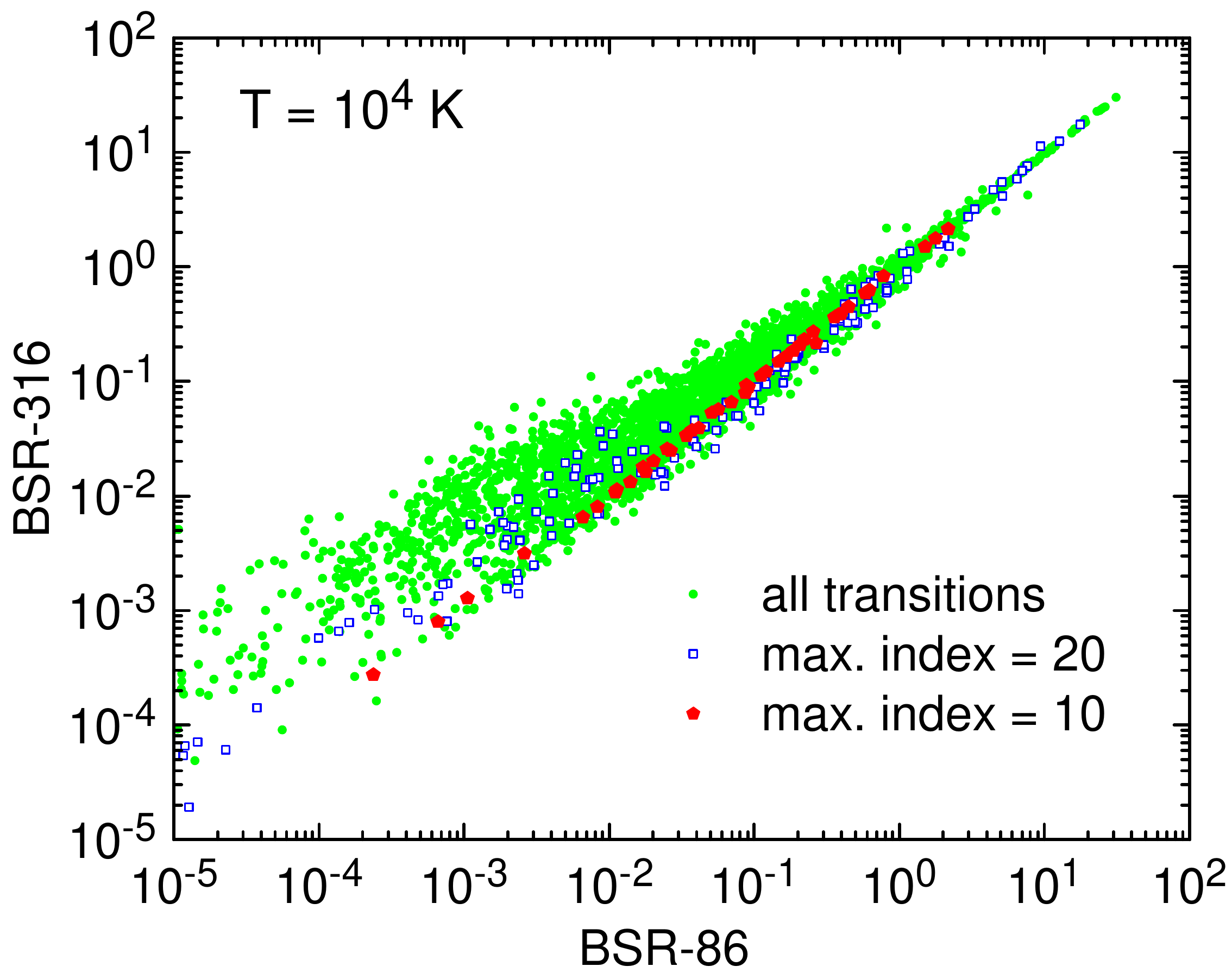}
\includegraphics[width=0.33\textwidth,clip]{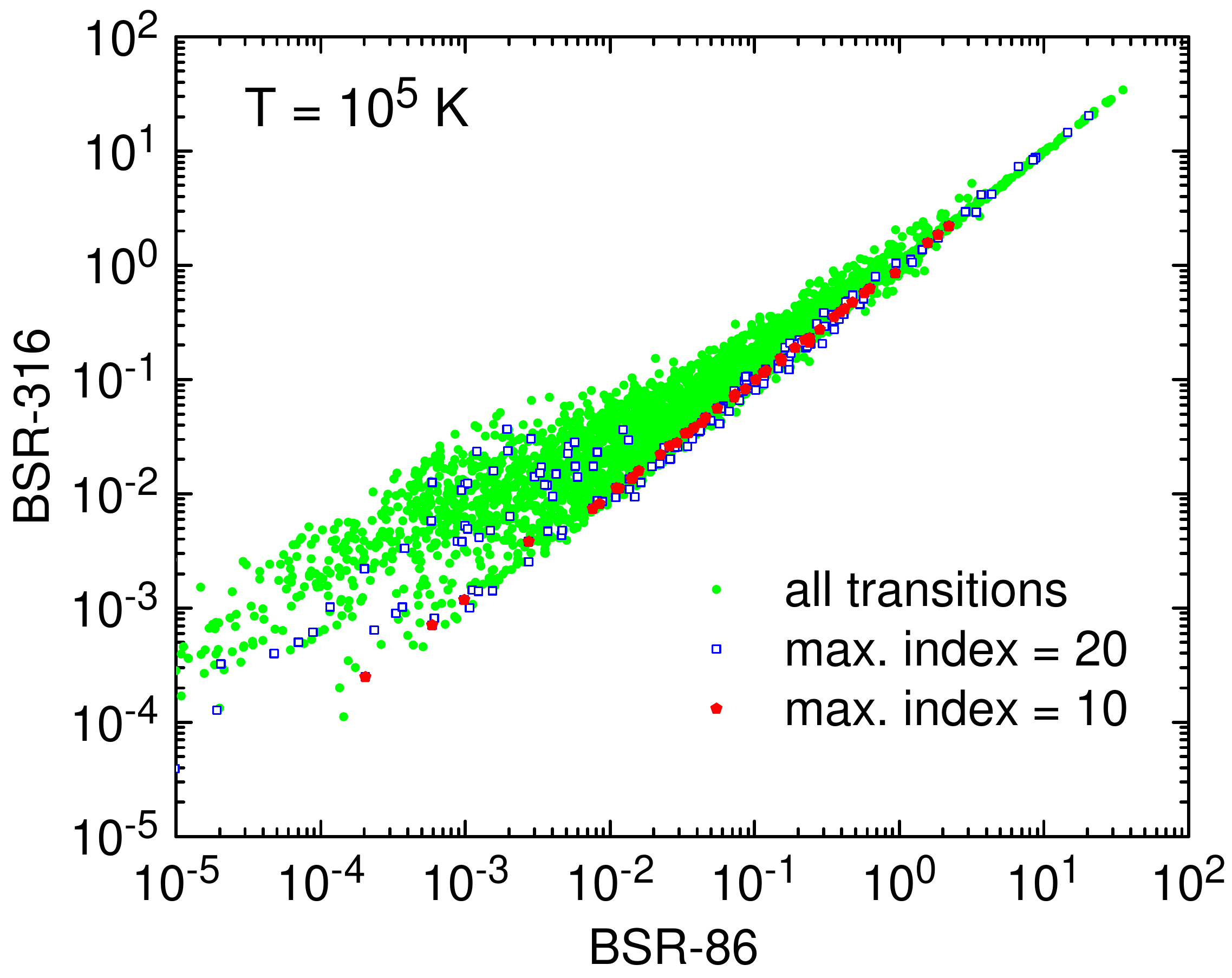}
\includegraphics[width=0.33\textwidth,clip]{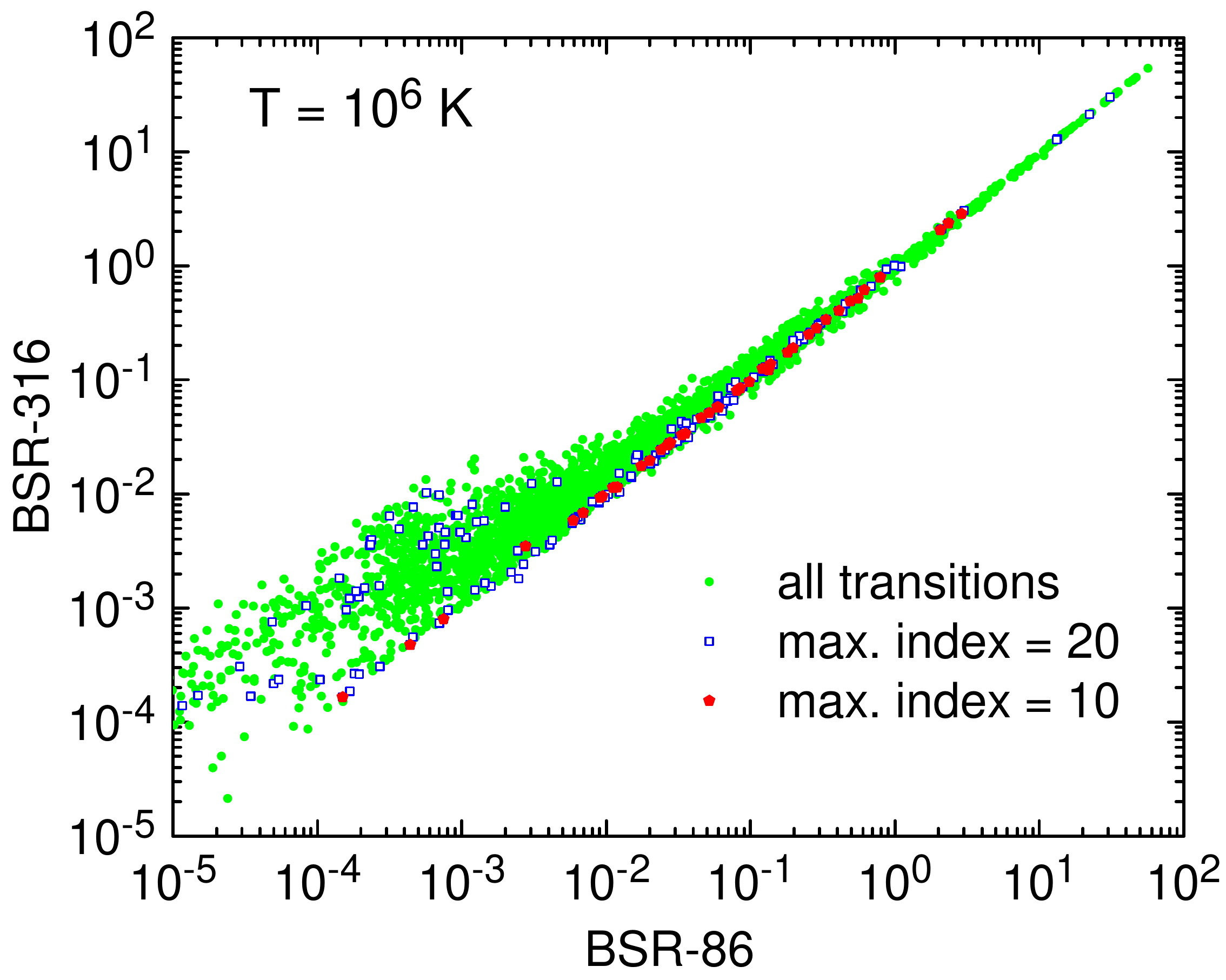}   
\caption{Comparison of effective collision strengths obtained in the present \hbox{BSR-86} and
          \hbox{BSR-316}  models. In addition to all transitions, we illustrate the situation for the
          lowest 10 and 20 states, respectively.
          $x$-axis: $\Upsilon$ results for a certain transition calculated with BSR-86 model;
          $y$-axis: $\Upsilon$ results for the same transition calculated with BSR-316 model.
         }
   \label{fig:COMP_BSR}
\end{figure*}

Figure~\ref{fig:omega} exhibits the collision strengths for different models at the electron
energy of $E=20\,\mathrm{Ry}$, i.e., above the resonance region.
This allows us to compare the background collision strengths. 
We found close agreement between the \hbox{BSR-86} and \hbox{BSR-316} background~$\Omega$ for the majority of transitions. 
The differences with the distorted-wave (DW) results of \cite{bhatia2006} reflect the close-coupling effects
and, once again, the different target descriptions.
Somewhat surprisingly, however, we notice much larger discrepancies with the \hbox{RM-37} than with the DW results, 
even though the former model includes close-coupling and
employs relatively accurate target wave\-functions.
Such large deviations of the \hbox{RM-37} collision strengths from both the \hbox{BSR-86} and DW results again
suggests the presence of pseudo\-resonances for some transitions.
This conclusion is supported when plotting the \hbox{RM-37} data
presented in their online tables.
For the problematic transitions, i.e., those far away from the diagonal line in
Fig.~\ref{fig:omega}, the \hbox{RM-37} collision strengths exhibit an unphysical energy dependence,
with broad resonances at higher energies.
This is indeed a typical situation for \hbox{R-matrix} calculations with the standard Belfast codes,
unless special care is taken to avoid these pseudo\-resonances, as outlined by~\cite{gorczyca1997}. 

In fact, the lack of balance between the \hbox{$N$-electron} (target) plus projectile scattering part 
and the \hbox{$(N+1)$-electron} pure
bound part of the CC expansion~(\ref{eq:RM}) may also have caused the appearance of pseudo\-resonance structure
in the \hbox{DARC-86} calculations. Their target CI expansions and the $(N+1)$-electron $\chi_i$ functions
contain configurations with a 2s hole, whereas no scattering channels with excitations out of the 2s sub\-shell 
were included. This may have led to an overestimate of the 2s resonance contribution.       

Figure~\ref{fig:COMP_BSR} presents a comparison of effective collision strengths from 
the present \hbox{BSR-86} and \hbox{BSR-316} models.
As expected, the corrections for the strong transitions are small and the corresponding
rates are expected to be converged.
The rate coefficients for all
transitions between the lowest 10 levels with configurations $\mathrm{2s^2\,2p^4}$,
$\mathrm{2s\,2p^5}$, and $\mathrm{2p^6}$ are stable against changes in the size of the CC expansion,
for both strong dipole and
weak inter\-combination transitions.

However, some transitions to the first excited $\mathrm{2s^2\,2p^3\,3}l$ states already
exhibit noticeable changes in their effective collision strengths.
There is a typical enhancement due to additional resonances, which was previously pointed
out by Fern\'andez-Menchero~{\it et al.} for $\mathrm{Fe^{20+}}$~\cite{fernandez-menchero2016a}.
A more detailed analysis shows that the most dramatic changes occur for the weak two- and
three-electron transitions from the $\mathrm{2s\,2p^5}$ and $\mathrm{2p^6}$ levels
to the $\mathrm{2s^2\,2p^3\,3}l$ states.
In these cases, the enhancement of the collision rates is not only due to additional
resonance structure but also due to changes (mainly enhancement) of the background collision strengths.
The resonance structure for these transitions also changes dramatically due to the inclusion
of the $\mathrm{2s\,2p^3\,3}l$ target states that lead to the appearance of strong resonances
with configurations $\mathrm{2s\,2p^3}\,3l\,3l'$.

\section{\label{sec:summary} Summary and conclusions} 

In response to recent criticism of the BSR method, and in order to resolve
the reasons for the large discrepancies between existing datasets, we have presented
new extensive calculations of the oscillator strengths and effective collision strengths
for the $\mathrm{Mg}^{4+}$ ion. Significant attention was devoted to the uncertainties in the scattering data
due to the target structure description and the size of the close-coupling expansion.
Our detailed and comprehensive comparison of the existing data\-sets
allows us to draw conclusions about their likely accuracy.

First, we independently obtained electron-impact excitation collision strengths
for the $\mathrm{Mg}^{4+}$ ion with the same number of
target states in the close-coupling expansion as in the recent \hbox{DARC-86}  and  \hbox{BSR-86 (TS)}
calculations. Comparing the results directly reveals the influence of the different structure
descriptions. The good agreement between our oscillator strengths with those from extensive MCHF calculations
indicates that our target structure description is the most accurate used so far.
The significant differences in the collision strengths seen with the \hbox{DARC-86} results in many cases suggest that
the target structure is the main source of inaccuracy and uncertainty in these calculations,
with the largest effects on transitions involving highly excited levels.
There are also indications of a remaining pseudo\-resonance structure in the \hbox{DARC-86} calculations.

On the other hand, we obtained very reasonable agreement with the results from the previous 
\hbox{BSR-86 (TS)} calculation~\cite{tayal2015}, except for excitation of the very last level (\#86) at low temperatures.
For this particular state, the deviations could simply be explained by too coarse of an energy grid
in the near-threshold region of the \hbox{BSR-86 (TS)} calculation.
Our findings, therefore, contradict the conclusions of Aggarwal and Keenan~\cite{aggarwal2016b} regarding a general inaccuracy
of the \hbox{BSR-86 (TS)} results.  Rather than concentrating on comparing collision strengths
for a single level (for which we explained the apparent problem),
our conclusions are based on a comprehensive comparison for all transitions.

We also  compared the present results with other available data\-sets. Overall good agreement
was obtained with the DW collision strengths of \cite{bhatia2006} at electron energies
above the resonance region.
The noticeable difference for some transitions is again connected with the too limited CI expansions
in the DW calculations. The largest discrepancies were seen with the \hbox{RM-37} 
results~\cite{hudson2009b}. The latter calculations are likely the least accurate
due to several reasons, including a still possibly insufficient quality of the target description,
a smaller CC expansion, and the indication of pseudo\-resonances.

Finally, we carried out much more extensive \hbox{BSR-316}  calculations, which additionally
included the  $\mathrm{2s^2\,2p^4\,4}l$ and $\mathrm{2s\,2p^5\,3}l$ target states.
This led to the appearance of new resonances.  Also, significant corrections were revealed
for transitions to high-lying states and between closely-spaced levels.
Due to the superior target structure generated in the present work and the larger CC expansions,
we believe that the present results are the currently best for
electron-impact excitation of $\mathrm{Mg}^{4+}$.
The differences between the \hbox{BSR-86} and the \hbox{BSR-316} results
may in fact serve as an uncertainty estimate for the available
excitation rates. Our final table contains the radiative and collision parameters
for transitions between all 316 target states. We are most confident
regarding the accuracy of the collision strengths between the first 86 states
presented in \hbox{Table~\ref{tab:energies}}.
The higher-lying states were included mainly to check the convergence
of the results.  Improving the accuracy of the collision strengths for these states
would require even larger calculations whose computational demands are
beyond currently available resources.

\section*{Acknowledgements}
We thank Dr.~S.~S.~Tayal sending data (including unpublished collision strengths)
in electronic form.
This work was supported by the United States National Science Foundation
through grants No.~PHY-1403245 and No.~PHY-1520970.  The numerical calculations were performed on \hbox{STAMPEDE}
at the Texas Advanced Computing Center.  They were made possible by the XSEDE allocation
No.~PHY-090031. K.~W.\ was sponsored by the China
Scholarship Council and would like to thank Drake University for the hospitality during his visit.

\bibliographystyle{apsrev4-1}

\begin{thebibliography}{20}%
\makeatletter
\providecommand \@ifxundefined [1]{%
 \@ifx{#1\undefined}
}%
\providecommand \@ifnum [1]{%
 \ifnum #1\expandafter \@firstoftwo
 \else \expandafter \@secondoftwo
 \fi
}%
\providecommand \@ifx [1]{%
 \ifx #1\expandafter \@firstoftwo
 \else \expandafter \@secondoftwo
 \fi
}%
\providecommand \natexlab [1]{#1}%
\providecommand \enquote  [1]{``#1''}%
\providecommand \bibnamefont  [1]{#1}%
\providecommand \bibfnamefont [1]{#1}%
\providecommand \citenamefont [1]{#1}%
\providecommand \href@noop [0]{\@secondoftwo}%
\providecommand \href [0]{\begingroup \@sanitize@url \@href}%
\providecommand \@href[1]{\@@startlink{#1}\@@href}%
\providecommand \@@href[1]{\endgroup#1\@@endlink}%
\providecommand \@sanitize@url [0]{\catcode `\\12\catcode `\$12\catcode
  `\&12\catcode `\#12\catcode `\^12\catcode `\_12\catcode `\%12\relax}%
\providecommand \@@startlink[1]{}%
\providecommand \@@endlink[0]{}%
\providecommand \url  [0]{\begingroup\@sanitize@url \@url }%
\providecommand \@url [1]{\endgroup\@href {#1}{\urlprefix }}%
\providecommand \urlprefix  [0]{URL }%
\providecommand \Eprint [0]{\href }%
\providecommand \doibase [0]{http://dx.doi.org/}%
\providecommand \selectlanguage [0]{\@gobble}%
\providecommand \bibinfo  [0]{\@secondoftwo}%
\providecommand \bibfield  [0]{\@secondoftwo}%
\providecommand \translation [1]{[#1]}%
\providecommand \BibitemOpen [0]{}%
\providecommand \bibitemStop [0]{}%
\providecommand \bibitemNoStop [0]{.\EOS\space}%
\providecommand \EOS [0]{\spacefactor3000\relax}%
\providecommand \BibitemShut  [1]{\csname bibitem#1\endcsname}%
\let\auto@bib@innerbib\@empty
\bibitem [{\citenamefont {Hudson}\ \emph {et~al.}(2009)\citenamefont {Hudson},
  \citenamefont {Ramsbottom}, \citenamefont {Norrington},\ and\ \citenamefont
  {Scott}}]{hudson2009b}%
  \BibitemOpen
  \bibfield  {author} {\bibinfo {author} {\bibfnamefont {C.~E.}\ \bibnamefont
  {Hudson}}, \bibinfo {author} {\bibfnamefont {C.~A.}\ \bibnamefont
  {Ramsbottom}}, \bibinfo {author} {\bibfnamefont {P.~H.}\ \bibnamefont
  {Norrington}}, \ and\ \bibinfo {author} {\bibfnamefont {M.~P.}\ \bibnamefont
  {Scott}},\ }\href {\doibase 10.1051/0004-6361:200809788} {\bibfield
  {journal} {\bibinfo  {journal} {Astron. Astroph.}\ }\textbf {\bibinfo
  {volume} {494}},\ \bibinfo {pages} {729} (\bibinfo {year}
  {2009})}\BibitemShut {NoStop}%
\bibitem [{\citenamefont {Tayal}\ and\ \citenamefont
  {Sossah}(2015)}]{tayal2015}%
  \BibitemOpen
  \bibfield  {author} {\bibinfo {author} {\bibfnamefont {S.~S.}\ \bibnamefont
  {Tayal}}\ and\ \bibinfo {author} {\bibfnamefont {A.~M.}\ \bibnamefont
  {Sossah}},\ }\href {\doibase 10.1051/0004-6361/201425135} {\bibfield
  {journal} {\bibinfo  {journal} {Astron. Astroph.}\ }\textbf {\bibinfo
  {volume} {574}},\ \bibinfo {pages} {A87} (\bibinfo {year}
  {2015})}\BibitemShut {NoStop}%
\bibitem [{\citenamefont {Aggarwal}\ and\ \citenamefont
  {Keenan}(2017)}]{aggarwal2016b}%
  \BibitemOpen
  \bibfield  {author} {\bibinfo {author} {\bibfnamefont {K.~M.}\ \bibnamefont
  {Aggarwal}}\ and\ \bibinfo {author} {\bibfnamefont {F.~P.}\ \bibnamefont
  {Keenan}},\ }\href {\doibase 10.1139/cjp-2016-0490} {\bibfield  {journal}
  {\bibinfo  {journal} {Can. J. Phys.}\ }\textbf {\bibinfo {volume} {95}},\
  \bibinfo {pages} {9} (\bibinfo {year} {2017})}\BibitemShut {NoStop}%
\bibitem [{\citenamefont {Hibbert}(1975)}]{hibbert1975}%
  \BibitemOpen
  \bibfield  {author} {\bibinfo {author} {\bibfnamefont {A.}~\bibnamefont
  {Hibbert}},\ }\href {\doibase http://dx.doi.org/10.1016/0010-4655(75)90103-4}
  {\bibfield  {journal} {\bibinfo  {journal} {Comp. Phys. Comm.}\ }\textbf
  {\bibinfo {volume} {9}},\ \bibinfo {pages} {141 } (\bibinfo {year}
  {1975})}\BibitemShut {NoStop}%
\bibitem [{\citenamefont {Berrington}\ \emph {et~al.}(1995)\citenamefont
  {Berrington}, \citenamefont {Eissner},\ and\ \citenamefont
  {Norrington}}]{berrington1995}%
  \BibitemOpen
  \bibfield  {author} {\bibinfo {author} {\bibfnamefont {K.~A.}\ \bibnamefont
  {Berrington}}, \bibinfo {author} {\bibfnamefont {W.~B.}\ \bibnamefont
  {Eissner}}, \ and\ \bibinfo {author} {\bibfnamefont {P.~H.}\ \bibnamefont
  {Norrington}},\ }\href@noop {} {\bibfield  {journal} {\bibinfo  {journal}
  {Comp. Phys. Comm.}\ }\textbf {\bibinfo {volume} {92}},\ \bibinfo {pages}
  {290} (\bibinfo {year} {1995})}\BibitemShut {NoStop}%
\bibitem [{\citenamefont {Griffin}\ \emph {et~al.}(1998)\citenamefont
  {Griffin}, \citenamefont {Badnell},\ and\ \citenamefont
  {Pindzola}}]{griffin1998}%
  \BibitemOpen
  \bibfield  {author} {\bibinfo {author} {\bibfnamefont {D.~C.}\ \bibnamefont
  {Griffin}}, \bibinfo {author} {\bibfnamefont {N.~R.}\ \bibnamefont
  {Badnell}}, \ and\ \bibinfo {author} {\bibfnamefont {M.~S.}\ \bibnamefont
  {Pindzola}},\ }\href {\doibase 10.1088/0953-4075/31/16/022} {\bibfield
  {journal} {\bibinfo  {journal} {J. Phys. B: At. Mol. Opt. Phys.}\ }\textbf
  {\bibinfo {volume} {31}},\ \bibinfo {pages} {3713} (\bibinfo {year}
  {1998})}\BibitemShut {NoStop}%
\bibitem [{\citenamefont {Froese-Fischer}\ \emph {et~al.}(2007)\citenamefont
  {Froese-Fischer}, \citenamefont {Tachiev}, \citenamefont {Gaigalas},\ and\
  \citenamefont {Godefroid}}]{froese-fischer2007}%
  \BibitemOpen
  \bibfield  {author} {\bibinfo {author} {\bibfnamefont {C.}~\bibnamefont
  {Froese-Fischer}}, \bibinfo {author} {\bibfnamefont {G.}~\bibnamefont
  {Tachiev}}, \bibinfo {author} {\bibfnamefont {G.}~\bibnamefont {Gaigalas}}, \
  and\ \bibinfo {author} {\bibfnamefont {M.~R.}\ \bibnamefont {Godefroid}},\
  }\href {\doibase http://dx.doi.org/10.1016/j.cpc.2007.01.006} {\bibfield
  {journal} {\bibinfo  {journal} {Comp. Phys. Comm.}\ }\textbf {\bibinfo
  {volume} {176}},\ \bibinfo {pages} {559 } (\bibinfo {year}
  {2007})}\BibitemShut {NoStop}%
\bibitem [{\citenamefont {Zatsarinny}\ and\ \citenamefont
  {Bartschat}(2013)}]{zatsarinny2013}%
  \BibitemOpen
  \bibfield  {author} {\bibinfo {author} {\bibfnamefont {O.}~\bibnamefont
  {Zatsarinny}}\ and\ \bibinfo {author} {\bibfnamefont {K.}~\bibnamefont
  {Bartschat}},\ }\href@noop {} {\bibfield  {journal} {\bibinfo  {journal} {J.
  Phys. B: At. Mol. Opt. Phys.}\ }\textbf {\bibinfo {volume} {46}},\ \bibinfo
  {pages} {112001} (\bibinfo {year} {2013})}\BibitemShut {NoStop}%
\bibitem [{\citenamefont {Zatsarinny}(2006)}]{zatsarinny2006}%
  \BibitemOpen
  \bibfield  {author} {\bibinfo {author} {\bibfnamefont {O.}~\bibnamefont
  {Zatsarinny}},\ }\href {\doibase 10.1016/j.cpc.2005.10.006} {\bibfield
  {journal} {\bibinfo  {journal} {Comp. Phys. Comm.}\ }\textbf {\bibinfo
  {volume} {174}},\ \bibinfo {pages} {273} (\bibinfo {year}
  {2006})}\BibitemShut {NoStop}%
\bibitem [{\citenamefont {{Hudson}}\ and\ \citenamefont
  {{Bell}}(2006)}]{hudson2006}%
  \BibitemOpen
  \bibfield  {author} {\bibinfo {author} {\bibfnamefont {C.~E.}\ \bibnamefont
  {{Hudson}}}\ and\ \bibinfo {author} {\bibfnamefont {K.~L.}\ \bibnamefont
  {{Bell}}},\ }\href {\doibase 10.1051/0004-6361:20054744} {\bibfield
  {journal} {\bibinfo  {journal} {Astron. Astroph.}\ }\textbf {\bibinfo
  {volume} {452}},\ \bibinfo {pages} {1113} (\bibinfo {year}
  {2006})}\BibitemShut {NoStop}%
\bibitem [{\citenamefont {Editors}(2011)}]{PhysRevA.83.040001}%
  \BibitemOpen
  \bibfield  {author} {\bibinfo {author} {\bibnamefont {Editors}},\ }\href
  {\doibase 10.1103/PhysRevA.83.040001} {\bibfield  {journal} {\bibinfo
  {journal} {Phys. Rev. A}\ }\textbf {\bibinfo {volume} {83}},\ \bibinfo
  {pages} {040001} (\bibinfo {year} {2011})}\BibitemShut {NoStop}%
\bibitem [{\citenamefont {Chung}\ \emph {et~al.}(2016)\citenamefont {Chung},
  \citenamefont {Braams}, \citenamefont {Bartschat}, \citenamefont {Császár},
  \citenamefont {Drake}, \citenamefont {Kirchner}, \citenamefont {Kokoouline},\
  and\ \citenamefont {Tennyson}}]{chung2016}%
  \BibitemOpen
  \bibfield  {author} {\bibinfo {author} {\bibfnamefont {H.-K.}\ \bibnamefont
  {Chung}}, \bibinfo {author} {\bibfnamefont {B.~J.}\ \bibnamefont {Braams}},
  \bibinfo {author} {\bibfnamefont {K.}~\bibnamefont {Bartschat}}, \bibinfo
  {author} {\bibfnamefont {A.~G.}\ \bibnamefont {Császár}}, \bibinfo {author}
  {\bibfnamefont {G.~W.~F.}\ \bibnamefont {Drake}}, \bibinfo {author}
  {\bibfnamefont {T.}~\bibnamefont {Kirchner}}, \bibinfo {author}
  {\bibfnamefont {V.}~\bibnamefont {Kokoouline}}, \ and\ \bibinfo {author}
  {\bibfnamefont {J.}~\bibnamefont {Tennyson}},\ }\href {\doibase
  http://iopscience.iop.org/0022-3727/49/36/363002} {\bibfield  {journal}
  {\bibinfo  {journal} {J. Phys. D: Appl. Phys.}\ }\textbf {\bibinfo {volume}
  {49}},\ \bibinfo {pages} {363002} (\bibinfo {year} {2016})}\BibitemShut
  {NoStop}%
\bibitem [{\citenamefont {Mazzotta}\ \emph {et~al.}(1998)\citenamefont
  {Mazzotta}, \citenamefont {Mazzitelli}, \citenamefont {Colafrancesco},\ and\
  \citenamefont {Vittorio}}]{mazzotta1998}%
  \BibitemOpen
  \bibfield  {author} {\bibinfo {author} {\bibfnamefont {P.}~\bibnamefont
  {Mazzotta}}, \bibinfo {author} {\bibfnamefont {G.}~\bibnamefont
  {Mazzitelli}}, \bibinfo {author} {\bibfnamefont {S.}~\bibnamefont
  {Colafrancesco}}, \ and\ \bibinfo {author} {\bibfnamefont {N.}~\bibnamefont
  {Vittorio}},\ }\href {\doibase 10.1051/aas:1998330} {\bibfield  {journal}
  {\bibinfo  {journal} {Astron. Astroph. Suppl. Ser.}\ }\textbf {\bibinfo
  {volume} {133}},\ \bibinfo {pages} {403} (\bibinfo {year}
  {1998})}\BibitemShut {NoStop}%
\bibitem [{\citenamefont {Zatsarinny}\ and\ \citenamefont
  {Froese-Fischer}(2009)}]{zatsarinny2009}%
  \BibitemOpen
  \bibfield  {author} {\bibinfo {author} {\bibfnamefont {O.}~\bibnamefont
  {Zatsarinny}}\ and\ \bibinfo {author} {\bibfnamefont {C.}~\bibnamefont
  {Froese-Fischer}},\ }\href {\doibase
  http://dx.doi.org/10.1016/j.cpc.2009.06.007} {\bibfield  {journal} {\bibinfo
  {journal} {Comp. Phys. Comm.}\ }\textbf {\bibinfo {volume} {180}},\ \bibinfo
  {pages} {2041} (\bibinfo {year} {2009})}\BibitemShut {NoStop}%
\bibitem [{\citenamefont {Kramida}\ \emph {et~al.}(2016)\citenamefont
  {Kramida}, \citenamefont {{Yu.~Ralchenko}}, \citenamefont {Reader},\ and\
  \citenamefont {{and NIST ASD Team}}}]{nist2016}%
  \BibitemOpen
  \bibfield  {author} {\bibinfo {author} {\bibfnamefont {A.}~\bibnamefont
  {Kramida}}, \bibinfo {author} {\bibnamefont {{Yu.~Ralchenko}}}, \bibinfo
  {author} {\bibfnamefont {J.}~\bibnamefont {Reader}}, \ and\ \bibinfo {author}
  {\bibnamefont {{and NIST ASD Team}}},\ }\href@noop {} {}\bibinfo
  {howpublished} {{NIST Atomic Spectra Database (ver. 5.4), [Online].
  Available: {\tt{http://physics.nist.gov/asd}} [2016, September 30]. National
  Institute of Standards and Technology, Gaithersburg, MD.}} (\bibinfo {year}
  {2016})\BibitemShut {NoStop}%
\bibitem [{\citenamefont {Froese-Fischer}\ and\ \citenamefont
  {Tachiev}(2004)}]{froese-fischer2004}%
  \BibitemOpen
  \bibfield  {author} {\bibinfo {author} {\bibfnamefont {C.}~\bibnamefont
  {Froese-Fischer}}\ and\ \bibinfo {author} {\bibfnamefont {G.}~\bibnamefont
  {Tachiev}},\ }\href {\doibase http://dx.doi.org/10.1016/j.adt.2004.02.001}
  {\bibfield  {journal} {\bibinfo  {journal} {At. Data Nucl. Data Tables\/}\
  }\textbf {\bibinfo {volume} {87}},\ \bibinfo {pages} {1} (\bibinfo {year}
  {2004})}\BibitemShut {NoStop}%
\bibitem [{\citenamefont {Burke}(2011)}]{burke2011}%
  \BibitemOpen
  \bibfield  {author} {\bibinfo {author} {\bibfnamefont {P.~G.}\ \bibnamefont
  {Burke}},\ }\href@noop {} {\emph {\bibinfo {title} {$R$-Matrix of Atomic
  Collisions: Application to Atomic, Molecular, and Optical Processes}}}\
  (\bibinfo  {publisher} {Springer-Verlag, New-York},\ \bibinfo {year}
  {2011})\BibitemShut {NoStop}%
\bibitem [{\citenamefont {Bhatia}\ \emph {et~al.}(2006)\citenamefont {Bhatia},
  \citenamefont {Landi},\ and\ \citenamefont {Eissner}}]{bhatia2006}%
  \BibitemOpen
  \bibfield  {author} {\bibinfo {author} {\bibfnamefont {A.~K.}\ \bibnamefont
  {Bhatia}}, \bibinfo {author} {\bibfnamefont {E.}~\bibnamefont {Landi}}, \
  and\ \bibinfo {author} {\bibfnamefont {W.}~\bibnamefont {Eissner}},\ }\href
  {\doibase http://dx.doi.org/10.1016/j.adt.2005.09.002} {\bibfield  {journal}
  {\bibinfo  {journal} {At. Data Nucl. Data Tables\/}\ }\textbf {\bibinfo
  {volume} {92}},\ \bibinfo {pages} {105 } (\bibinfo {year}
  {2006})}\BibitemShut {NoStop}%
\bibitem [{\citenamefont {Gorczyca}\ and\ \citenamefont
  {Badnell}(1997)}]{gorczyca1997}%
  \BibitemOpen
  \bibfield  {author} {\bibinfo {author} {\bibfnamefont {T.~W.}\ \bibnamefont
  {Gorczyca}}\ and\ \bibinfo {author} {\bibfnamefont {N.~R.}\ \bibnamefont
  {Badnell}},\ }\href {\doibase http://iopscience.iop.org/0953-4075/30/17/017}
  {\bibfield  {journal} {\bibinfo  {journal} {J. Phys. B: At. Mol. Opt. Phys.}\
  }\textbf {\bibinfo {volume} {30}},\ \bibinfo {pages} {3897} (\bibinfo {year}
  {1997})}\BibitemShut {NoStop}%
\bibitem [{\citenamefont {Fern\'andez-Menchero}\ \emph
  {et~al.}(2016)\citenamefont {Fern\'andez-Menchero}, \citenamefont {Giunta},
  \citenamefont {{Del~Zanna}},\ and\ \citenamefont
  {Badnell}}]{fernandez-menchero2016a}%
  \BibitemOpen
  \bibfield  {author} {\bibinfo {author} {\bibfnamefont {L.}~\bibnamefont
  {Fern\'andez-Menchero}}, \bibinfo {author} {\bibfnamefont {A.~S.}\
  \bibnamefont {Giunta}}, \bibinfo {author} {\bibfnamefont {G.}~\bibnamefont
  {{Del~Zanna}}}, \ and\ \bibinfo {author} {\bibfnamefont {N.~R.}\ \bibnamefont
  {Badnell}},\ }\href {\doibase
  http://dx.doi.org/10.1088/0953-4075/49/8/085203} {\bibfield  {journal}
  {\bibinfo  {journal} {J. Phys. B: At. Mol. Opt. Phys.}\ }\textbf {\bibinfo
  {volume} {49}},\ \bibinfo {pages} {085203} (\bibinfo {year}
  {2016})}\BibitemShut {NoStop}%
\end{thebibliography}
%

\end{document}